\documentclass[twocolumn,showpacs,preprintnumbers,amsmath,amssymb,prb,
superscriptaddress]{revtex4-1}

\usepackage{graphicx}
\usepackage{dcolumn}
\usepackage{bm}
\usepackage{dcolumn}
\usepackage{amsmath}
\usepackage{amssymb}
\usepackage{longtable}
\usepackage{array}
\usepackage[version=3]{mhchem}
\usepackage{epic}
\usepackage{epstopdf}
\usepackage{verbatim}

\usepackage[usenames,dvipsnames,svgnames,table]{xcolor}

\newcolumntype{C}[1]{>{\centering}m{#1}}
\newcommand{\etal}{\textit{et~al.}}
\newcommand{\bv}[1]{{\boldsymbol #1}}

\newcommand{\muB}{\mu _\text{B}}

\begin{document}

\title{Theoretical study on the anisotropic electronic structure of antiferromagnetic \ce{BaFe2As2} and Co-doped \boldmath \ce{Ba(Fe_{1-$x$}Co_{$x$})2As2} \unboldmath as seen by angle-resolved photoemission} 
\author{Gerald Derondeau}    \email{gerald.derondeau@cup.uni-muenchen.de}
\affiliation{%
  Department  Chemie,  Physikalische  Chemie,  Universit\"at  M\"unchen,
  Butenandtstr.  5-13, 81377 M\"unchen, Germany\\}
\author{J\"urgen Braun}
\affiliation{%
  Department  Chemie,  Physikalische  Chemie,  Universit\"at  M\"unchen,
  Butenandtstr. 5-13, 81377 M\"unchen, Germany\\}
\author{Hubert Ebert}
\affiliation{%
  Department  Chemie,  Physikalische  Chemie,  Universit\"at  M\"unchen,
  Butenandtstr. 5-13, 81377 M\"unchen, Germany\\}
\author{J\'an Min\'ar}
\affiliation{%
  Department  Chemie,  Physikalische  Chemie,  Universit\"at  M\"unchen,
  Butenandtstr. 5-13, 81377 M\"unchen, Germany\\}
\affiliation{%
  NewTechnologies-Research Center, University of West Bohemia, Pilsen, Czech Republic\\}

\date{\today}


\begin{abstract}
By means of one-step model calculations the strong in-plane anisotropy seen in angle-resolved photoemission of the well-known iron pnictide prototype compounds \ce{BaFe2As2} and \ce{Ba(Fe_{1-$x$}Co_{$x$})2As2} in their low-temperature antiferromagnetic phases is investigated. The fully-relativistic calculations are based on the Korringa-Kohn-Rostoker-Green function approach combined with the coherent potential approximation alloy theory to account for the disorder induced by Co substitution on Fe sites in a reliable way. The results of the calculations can be compared directly to experimental spectra of detwinned single crystals. One finds very good agreement with experiment and can reveal all features of the electronic structure contributing to the in-plane anisotropy. In particular the local density approximation can capture most of the correlation effects for the investigated system without the need for more advanced techniques. In addition, the evolution of the anisotropy for increasing Co concentration $x$ in \ce{Ba(Fe_{1-$x$}Co_{$x$})2As2} can be tracked almost continuously. The results are also used to discuss surface effects and it is possible to identify clear signatures to conclude about different types of surface termination.

\end{abstract}
                
\maketitle


\section{Introduction}
Nowadays the family of iron pnictides is a well-established and important prototype system for unconventional high-temperature superconductivity. Starting with the first famous compound \ce{La(O_{1-$x$}F_$x$)FeAs} \cite{KWHH08,TIA+08} in 2008, today several different sub-families with a wide structural variety are known.

All different groups of iron pnictides share some common physical properties, such as their interesting and sometimes puzzling magnetic behavior. Most compounds show a phase transition at low temperatures from a tetragonal to an orthorhombic crystal symmetry which is typically accompanied by the formation of long-range antiferromagnetic order.\cite{SLS+09,YLH+08} It is common believe that the suppression of these phase transitions for example by chemical substitution is crucial for the emergence of unconventional superconductivity.\cite{Sin08a,MSJD08} Although it is obvious that an understanding of the magnetic fluctuations in the iron pnictides is mandatory to unveil the physics underlying the superconductivity, this task has proven to be more complex than anticipated.\cite{MJ09,MJB+08,MSJD08}

For example, there was discussion in the literature whether the magnetic moments are better described by an itinerant\cite{MJ09,MSJD08,YLAA09,OKZ+09,FTO+09} or a localized\cite{HYPS09,YZO+09,CLLR08} model and there is up to now no consensus concerning the role of correlation effects\cite{JM09, SBP+11,WCM+12,LYM+08}. Furthermore, the magnitude of the magnetic moments is difficult to reproduce within density functional theory (DFT) and it is known to be quite sensitive to computational parameters.\cite{MJB+08,SLS+09,RTJ+08,GLK+11,VFB+12} 

One of the most important experimental tools to get insight into the electronic structure of the iron pnictides is angle-resolved photoemission spectroscopy (ARPES). There are numerous publications on this topic, although it was shown that DFT calculations have typically problems to reproduce all features of the ARPES spectra correctly.\cite{ZIE+09,EIZ+09,EIZ+09a,WCM+12,EZK+14} This is often ascribed to strong correlation effects, although this question is still under discussion.\cite{YHK11,WCM+12,FFVJ12}

Another important difficulty which so far is often ignored is the connection between the magnetic phase of the iron pnictides and the resulting consequences for ARPES. This is due to the formation of twinned crystals during the phase transition from tetragonal to orthorhombic and it results in mixed magnetic domains which are orthogonal to each other. Macroscopic tools like ARPES or transport measurements can so only see the averaged information, while information on the anisotropy is lost.\cite{YLC+11,KOK+11} This is a huge drawback considering a comprehensive study of the electronic structure in the iron pnictides, as it is known that the in-plane anisotropy plays a significant role.\cite{CAL+10,CAP+10,ZAY+09} In experiment it is possible to effectively detwin the crystals by applying uniaxial stress during the measurement. This was already done successfully for the 122-prototype \ce{BaFe2As2} in the undoped and in the Co-doped case. However, such measurements are connected with several technical difficulties and consequently they are rarely done.\cite{YLC+11,KOK+11} Yet, to fully understand the electronic properties of the iron pnictide superconductors in a comprehensive way and to get a deeper insight concerning the influence of the in-plane anisotropy in the magnetic phase such studies are absolutely mandatory. Although there is nowadays experimental data on detwinned crystals showing clearly the anisotropy in the Fermi surface there is hardly any theoretical work focusing on this problem of magnetic anisotropy in ARPES data.

In this work this issue is addressed by a comprehensive DFT study on the magnetic phase of \ce{Ba(Fe_{1-$x$}Co_{$x$})2As2} and on the corresponding ARPES spectra. The computational results can be directly compared to the available experimental ARPES data on detwinned crystals.\cite{YLC+11,KOK+11} 

In order to deal with this complex situation the Korringa-Kohn-Rostoker-Green function (KKR-GF) approach is used, which was already shown to be indeed a very useful and accurate tool to deal with the iron pnictides.\cite{DPM+14} The impact of disorder due to substitution is dealt with by means of the coherent potential approximation (CPA), giving results fully compatible to supercell calculations and more reliable than those based on the virtual crystal approximation (VCA).\cite{DPM+14}

\section{Computational details}
All calculations have been performed self-consistently and fully relativistically within the four component Dirac formalism, using the Munich SPR-KKR program package.\cite{EKM11,SPR-KKR6.3_2}

The orthorhombic, antiferromagnetic phase of \ce{BaFe2As2} is investigated in its experimentally observed stripe spin state using a full 4-Fe unit cell. This implies antiferromagnetic chains along the $a$- and $c$-axes and ferromagnetic chains along the $b$-axis. The lattice parameters where chosen according to experimental X-ray data and the experimental As position $z$.\cite{RTJ+08} To account for the influence of substitution in \ce{Ba(Fe_{1-$x$}Co_{$x$})2As2} a linear interpolation for the lattice parameters with respect to the concentration $x$ is used based on available experimental data\cite{RTJ+08,SJM+08} and Vegard's law\cite{Veg21}. More details on the procedure can be found in a previous publication.\cite{DPM+14} The treatment of disorder introduced by substitution is dealt with by means of the CPA. The basis set considered for a $l_\text{max}=4$ including $s$, $p$, $d$, $f$ and $g$ orbitals. For the electronic structure calculations the local density approximation (LDA) exchange-correlation potential with the parameterization given by Vosko, Wilk and Nusair was applied.\cite{VWN80} 

The spectroscopical analysis is based on the fully relativistic one-step model of photoemission in its spin density matrix formulation. For more technical details on these calculations see Ref.\cite{MBE13,BMK+14}. The geometry of the spectroscopy setup was taken from experiment including a tilt of the sample around either the $a$ or $b$ axis. The incident light hit the sample under a constant polar angle $\theta_\text{light}=-45^\circ$ and an azimuthal angle $\phi_\text{light}$ of either $180^\circ$ or $270^\circ$. These geometries are referred to as $\bv q_{||\text{AFM}}$ and $\bv q_{||\text{FM}}$, meaning the direction of the incident light is either parallel to the antiferromagnetic or the ferromagnetic in-plane directions. The corresponding electrons were collected with an angle $\phi_\text{electron}$ of $0^\circ$ or $90^\circ$ and a varying angle $\theta_\text{electron}$ between $-15^\circ$ and $15^\circ$. This geometry is in line to the experimental setup. If not indicated otherwise, an As-terminated surface was chosen. However, the question of surface termination will be discussed in more detail in the following. 
%
\section{Results and Discussion}

\subsection{Magnetic moments}
To describe the anisotropy of the iron pnictides in ARPES calculations reasonably well one needs first to ensure that the spin-dependent potentials from the self-consistent field (SCF) calculations are accurate enough. Obviously, the magnetic ordering plays a significant role concerning the anisotropy of the electronic structure and hence the quality of the theoretical description of the ARPES spectra is determined by the quality of the spin-dependent potentials.

The most meaningful indication for a proper description of the magnetic state is good agreement with experimental data on the magnetic order. For the iron pnictides this is known to be a non-trivial task as the magnetic moments are often overestimated by DFT.\cite{SBP+11,MJB+08,MJ09} For the undoped mother compound \ce{BaFe2As2} a total magnetic moment of $0.73\muB$ was obtained. Experiment reports a total magnetic moment of approximately $0.9\muB$ from neutron diffraction\cite{HQB+08,SLS+09} while \ce{^{57}Fe}~M\"ossbauer spectroscopy\cite{RTJ+08,RTS+09} and $\mu$SR spectroscopy\cite{ABB+08,GAB+09} coherently give a value of around $0.5\muB$. Hence, the calculated total magnetic moment is found in good agreement with experiment and captures the proper order of magnitude accurately.\cite{SBP+11,MJB+08,MJ09,YLAA09,KOK+09,AC09}

\begin{figure}[b]
\vspace{-0.5cm}
\hspace{+1.5cm}\scalebox{0.70}{
\large
  \setlength{\unitlength}{0.0500bp}%
  \begin{picture}(7200.00,5040.00)%
      \put(814,801){\makebox(0,0)[r]{\strut{}0.0}}%
      \put(814,1770){\makebox(0,0)[r]{\strut{}0.2}}%
      \put(814,2740){\makebox(0,0)[r]{\strut{}0.4}}%
      \put(814,3709){\makebox(0,0)[r]{\strut{}0.6}}%
      \put(814,4678){\makebox(0,0)[r]{\strut{}0.8}}%
      \put(946,581){\makebox(0,0){\strut{} 0}}%
      \put(2330,581){\makebox(0,0){\strut{} 0.05}}%
      \put(3715,581){\makebox(0,0){\strut{} 0.1}}%
      \put(5099,581){\makebox(0,0){\strut{} 0.15}}%
      \put(5923,801){\makebox(0,0)[l]{\strut{}0.00}}%
      \put(5923,1770){\makebox(0,0)[l]{\strut{}0.02}}%
      \put(5923,2740){\makebox(0,0)[l]{\strut{}0.04}}%
      \put(5923,3709){\makebox(0,0)[l]{\strut{}0.06}}%
      \put(5923,4678){\makebox(0,0)[l]{\strut{}0.08}}%
      \put(176,2739){\rotatebox{-270}{\makebox(0,0){\strut{}\Large Spin moment $m_\text{s}$  $\left[ \mu _\text{B} \right]$}}}%
      \put(6692,2739){\rotatebox{-270}{\makebox(0,0){\strut{}\Large Orbital moment $m_\text{o}$  $\left[ \mu _\text{B} \right]$}}}%
      \put(3368,251){\makebox(0,0){\strut{}\Large  $x$ of \ce{Ba(Fe_{1-$x$}Co_$x$)_2As2}}}%
      \put(4176,4505){\makebox(0,0)[l]{\strut{}$m_\text{s}$(Fe)}}%
      \put(4176,4285){\makebox(0,0)[l]{\strut{}$m_\text{o}$(Fe)}}%
      \put(4176,4065){\makebox(0,0)[l]{\strut{}$m_\text{s}$(Co)}}%
      \put(4176,3845){\makebox(0,0)[l]{\strut{}$m_\text{o}$(Co)}}%
      \put(4176,3625){\makebox(0,0)[l]{\strut{}$m_\text{tot}$}}%
    \put(0,0){\includegraphics{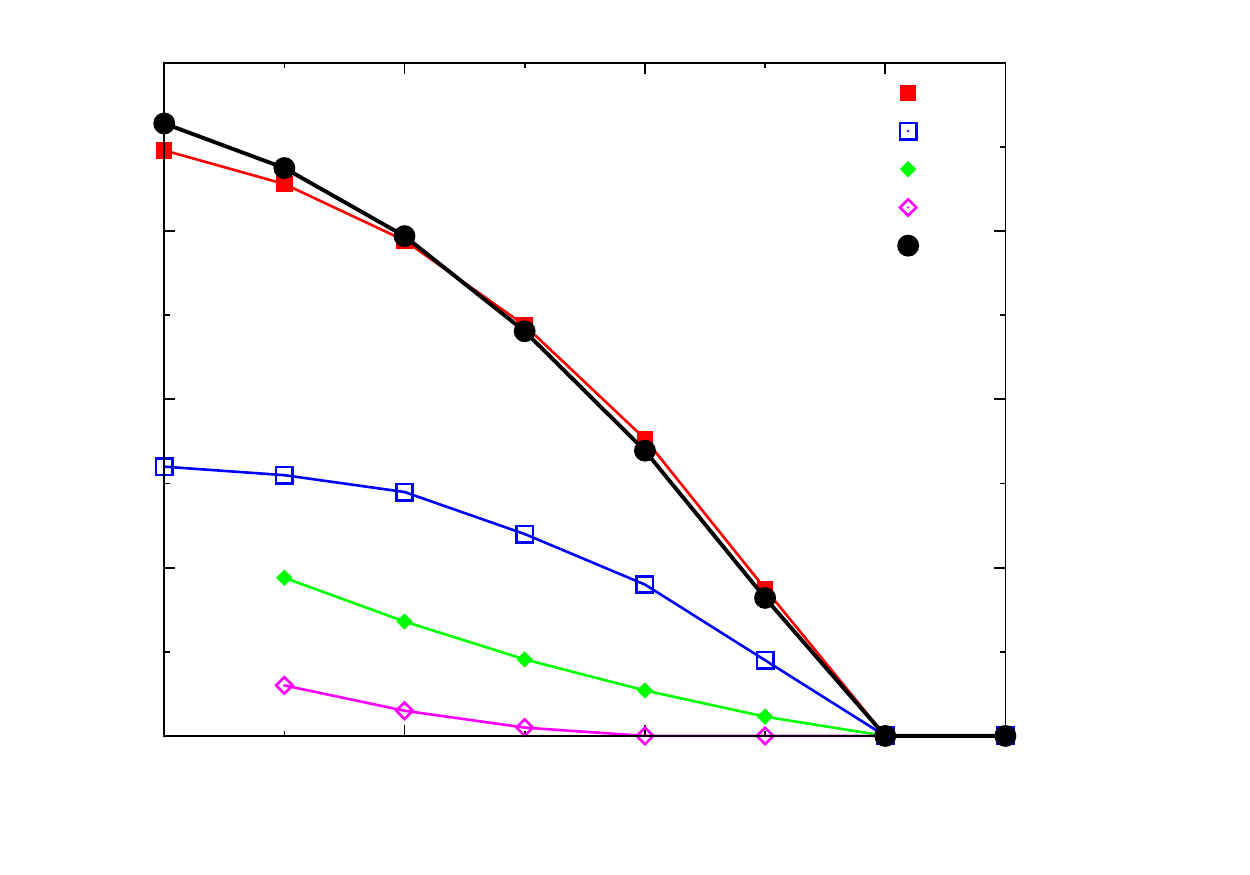}}%
  \end{picture}}
\vspace{-0.8cm}
\caption{(Color online) Magnetic spin moments ($m_\text{s}$) and orbital moments ($m_\text{o}$) of Fe and Co for increasing Co substitution $x$. The black line corresponds to the total (spin and orbital) magnetic moment which
 is the substitutionally averaged sum over all other contributions. The left axis corresponds to the spin magnetic moments $m_\text{s}$, while the right axis shows the orbital magnetic moments $m_\text{o}$.}\label{Fig_MagnMom}
\end{figure}
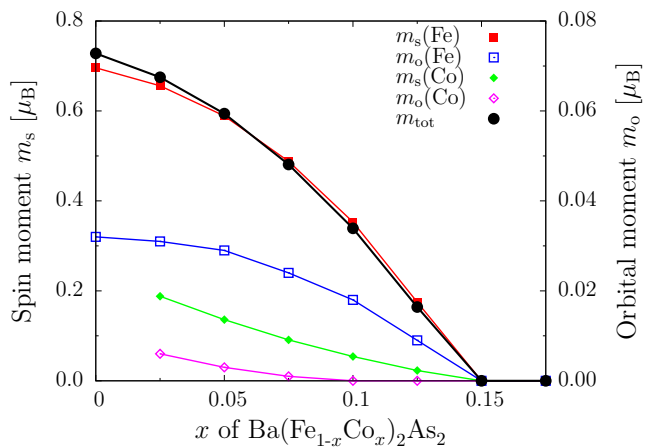

More importantly, the CPA allows to evaluate the substitution dependent self-consistent evolution of the magnetic moments with increasing Co concentration in \ce{Ba(Fe_{1-$x$}Co_{$x$})2As2}. The corresponding results are shown in Fig.~\ref{Fig_MagnMom}, where the results for spin and orbital magnetic moments are given in an atom-resolved way. The total magnetic moment is calculated as substitutionally averaged sum over all contributions.

In agreement with experiment the total magnetic moments shows a nearly linear decay until the long-range magnetic order disappears.\cite{LCA+09} In the calculations the critical Co substitution for the disappearance of antiferromagnetic order occurs for $x_\text{crit}=0.15$, which is in reasonably good agreement with the experimental value $x_\text{crit,exp}\approx 0.07$.\cite{CAK+09,KOK+09} It should be mentioned that the results in Fig.~\ref{Fig_MagnMom} are slightly improved with respect to experiment in comparison with our previous work\cite{DPM+14} due to the higher $l_\text{max}$ expansion used here. However, the trends in the magnetic moments and the resulting conclusions are the same.

\subsection{Anisotropy of the undoped compound}

As the one-step model of photoemission fully accounts for matrix-elements as well as for surface effects the resulting spectra can be directly compared to experimental ARPES data.

As stressed before, it is extremely difficult to see the magnetic anisotropy correctly in experimental spectra because of the twinning of crystals. Here reference is made especially to the work of Yi \etal \cite{YLC+11}, who did remarkable measurements on detwinned single crystals of \ce{BaFe2As2} and \ce{Ba(Fe_{1-$x$}Co_$x$)_2As2} by applying uniaxial stress to the crystals. Similar results were obtained for example by Kim \etal \cite{KOK+11}. In this context it is important to note that the Brillouin zone (BZ) of the magnetic 4-Fe spin-density-wave (SDW) state is only half the size compared to the BZ in the nonmagnetic 2-Fe state. For that reason it is most appropriate to use in the following the notation for the 4-Fe SDW~BZ where the information of $\bar \Gamma_\text{}$ and $\bar{\mathrm X_\text{}}$ from the nonmagnetic BZ is down-folded to one $\bar \Gamma$ point.\cite{YLC+11,YLA+09a}

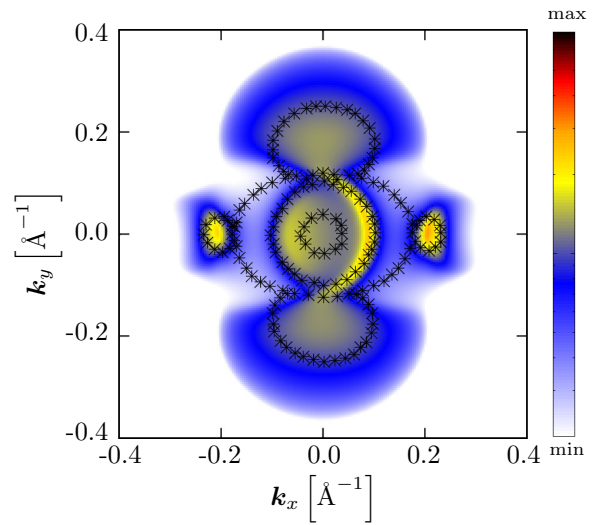
\begin{figure}[tb]
  \setlength{\unitlength}{0.68cm}\normalsize 
     \begin{picture}(12,10.0)
      \put(2.20,1.2){\includegraphics[height=8\unitlength,clip]{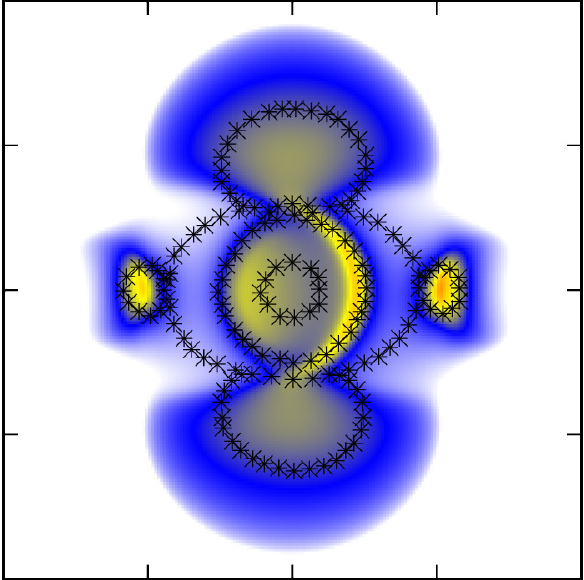}}
      \multiputlist(1.2,1.35)(0,1.95)[l]{-0.4,-0.2,~0.0,~0.2,~0.4}
      \multiputlist(1.85,0.9)(1.95,0)[l]{-0.4,-0.2,~0.0,~0.2,~0.4}
      \put(0.2,4.0){\rotatebox{90}{$\bv k_y \left[ \text{\AA}^{-1} \right]$}}
      \put(5.2,-0.1){{$\bv k_x \left[ \text{\AA}^{-1} \right]$}}
      \scriptsize
      \put(10.7,1.2){\includegraphics[height=8\unitlength,clip]{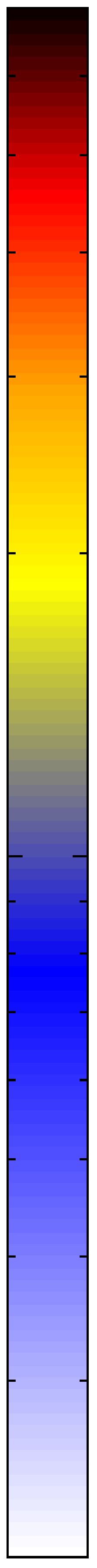}}
      \put(11.0,9.5){\makebox(0,0){max}}
      \put(11.0,1.0){\makebox(0,0){min}}
      \end{picture}
\caption{(Color online) Calculated ARPRES spectrum mapping the Fermi surface at $\bar \Gamma$ in the 4-Fe SDW BZ for a photon energy of $h\nu=22$\;eV. The overlay of black points is a reconstruction of the SDW BZ from experimental ARPES data, reproduced from the work of Yi~\etal \cite{YLC+11}.}\label{Fig_Cal_AFMain}
\end{figure}

In Fig.~\ref{Fig_Cal_AFMain} the Fermi surface around the $\bar \Gamma$ point is shown in the SDW BZ as calculated from the spin-dependent potentials for a photon energy of $h\nu=22$\;eV. The overlay of black points corresponds to the experimentally measured BZ, reproduced from the work of Yi~\etal \cite{YLC+11}. 
As can be seen, the agreement of the calculated Fermi surface and the experimental data is remarkably good. Characteristic are the bright intensity spots along the $\bv k_x$-direction (i.e. along $a$), corresponding directly to the antiferromagnetic order along the $a$-axis and the bigger pedal-like structures along the $\bv k_y$-direction (i.e. along $b$) which corresponds to the ferromagnetic order along the $b$-axis.

\begin{figure}[b]
  \setlength{\unitlength}{0.61cm}\small
     \begin{picture}(5.9,8)
      \put(1.10,1.2){\includegraphics[height=5.5\unitlength,clip]{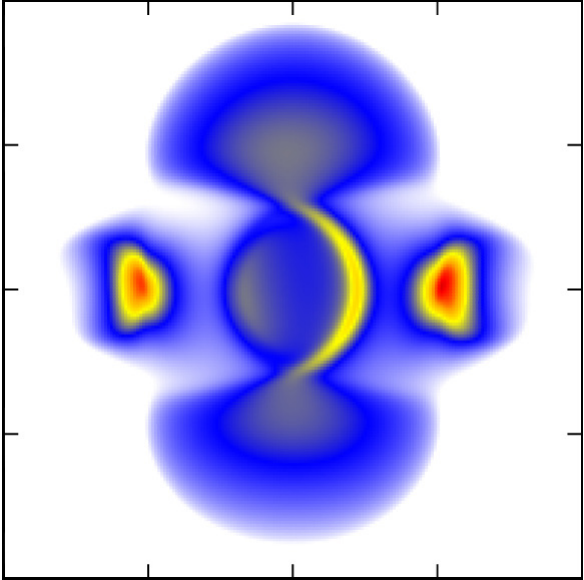}}
      \multiputlist(0.2,1.35)(0,1.3)[l]{-0.4,-0.2,~0.0,~0.2,~0.4}
      \multiputlist(0.75,0.9)(1.3,0)[l]{-0.4,-0.2,~0.0,~0.2,~0.4}
      \put(-0.6,3.0){\rotatebox{90}{$\bv k_y \left[ \text{\AA}^{-1} \right]$}}
      \put(2.7,-0.1){{$\bv k_x \left[ \text{\AA}^{-1} \right]$}}
      \put(0.65,7.0){\large  (a)}
      \put(2.75,7.0){\large $\bv q_{||\text{FM}}$}
      \end{picture}
     \begin{picture}(7.3,8)
      \put(1.20,1.2){\includegraphics[height=5.5\unitlength,clip]{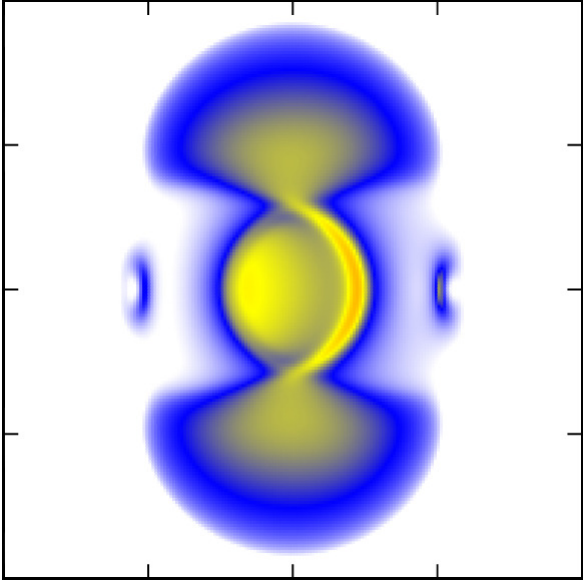}}
      \multiputlist(0.85,0.9)(1.3,0)[l]{-0.4,-0.2,~0.0,~0.2,~0.4}
      \put(2.8,-0.1){{$\bv k_x \left[ \text{\AA}^{-1} \right]$}}
      \put(0.65,6.95){\large (b)}
      \put(2.75,6.95){\large $\bv q_{||\text{AFM}}$}
      \scriptsize
      \put(7.0,1.2){\includegraphics[height=5.5\unitlength,clip]{figScale.pdf}}
      \put(7.3,6.95){\makebox(0,0){max}}
      \put(7.3,1.0){\makebox(0,0){min}}
      \end{picture}
\caption{(Color online) Calculated ARPES spectra mapping the Fermi surface for the polarization of light being either parallel to the ferromagnetic $b$-axis (a) or parallel to the antiferromagnetic $a$-axis (b). 
}\label{Fig_Cal_AFpol}
\end{figure}

It should be noted that in Fig.~\ref{Fig_Cal_AFMain} the intensity over two different light polarizations was averaged, namely for the direction of the incident light either parallel to the antiferromagnetic $a$-axis  $\bv q_{||\text{AFM}}$ or parallel to the ferromagnetic $b$-axis  $\bv q_{||\text{FM}}$. All features of the electronic structure are visible for both polarizations of light. However, the intensity patterns vary notably with the polarization due to matrix element effects, indicating strong multiorbital character, just as seen in experiment.\cite{YLC+11} If not indicated otherwise this averaging will be applied in the following. For comparison the two contributions to the total Fermi surface for $h\nu=22$\;eV are shown polarization-resolved in Fig.~\ref{Fig_Cal_AFpol}, for the incident light direction being either parallel to the $b$-axis $\bv q_{||\text{FM}}$\;(Fig.~\ref{Fig_Cal_AFpol} (a)) or parallel to the $a$-axis $\bv q_{||\text{AFM}}$\;(Fig.~\ref{Fig_Cal_AFpol} (b)).

It can be seen that for $\bv q_{||\text{FM}}$ the intensity of the bright spots along the $a$-axis is significantly enhanced while for $\bv q_{||\text{AFM}}$ the intensity around the inner circle of $\bar \Gamma$ is enhanced. This polarization dependence is again in full agreement with the experimental findings.\cite{YLC+11}

\begin{figure}[tb]
  \setlength{\unitlength}{0.61cm}\small
     \begin{picture}(5.9,8)
      \put(1.10,1.2){\includegraphics[height=5.5\unitlength,clip]{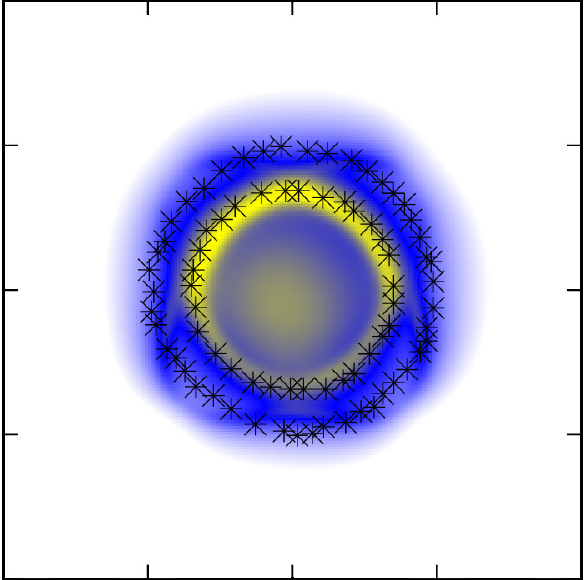}}
      \multiputlist(0.2,1.35)(0,1.3)[l]{-0.4,-0.2,~0.0,~0.2,~0.4}
      \multiputlist(0.75,0.9)(1.3,0)[l]{-0.4,-0.2,~0.0,~0.2,~0.4}
      \put(-0.6,3.0){\rotatebox{90}{$\bv k_y \left[ \text{\AA}^{-1} \right]$}}
      \put(2.7,-0.1){{$\bv k_x \left[ \text{\AA}^{-1} \right]$}}
      \put(0.65,7.0){\large  (a)}
      \put(2.1,7.0){\large nonmagnetic}
      \end{picture}
     \begin{picture}(7.3,8)
      \put(1.20,1.2){\includegraphics[height=5.5\unitlength,clip]{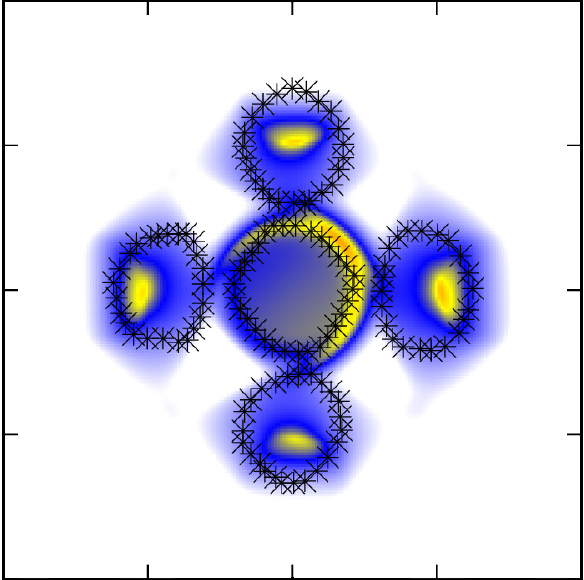}}
      \multiputlist(0.85,0.9)(1.3,0)[l]{-0.4,-0.2,~0.0,~0.2,~0.4}
      \put(2.8,-0.1){{$\bv k_x \left[ \text{\AA}^{-1} \right]$}}
      \put(0.65,6.95){\large (b)}
      \put(2.75,6.95){\large twinned}
      \scriptsize
      \put(7.0,1.2){\includegraphics[height=5.5\unitlength,clip]{figScale.pdf}}
      \put(7.3,6.95){\makebox(0,0){max}}
      \put(7.3,1.0){\makebox(0,0){min}}
      \end{picture}
\caption{(Color online) Calculated Fermi surfaces as seen by ARPES for the nonmagnetic phase (a) and a hypothetical twinned magnetic phase (b). The twinned calculation is based on a superposition of two antiferromagnetic cells which are rotated by $90^\circ$ against each other. The overlay of black points is reproduced from the experimental ARPES data of Yi \etal\cite{YLA+09a}}\label{Fig_Cal_Twin}
\end{figure}

At this point it was shown that the detwinned, antiferromagnetic Fermi surface obtained by the calculations agrees very well with experiment. One may also ask how the Fermi surface of a twinned, antiferromagnetic crystal should look like and how does it differ from the nonmagnetic case. Therefore, the calculated Fermi surfaces for both cases are shown in Fig.~\ref{Fig_Cal_Twin} and compared with experimental ARPES data\cite{YLA+09a} of twinned \ce{BaFe2As2} crystals at $T=150$\;K (a) and $T=10$\;K (b). Please note that the transition from a paramagnetic to an antiferromagnetic state occurs at around $T=140$\;K, accordingly the experimental data shown as overlay of black points corresponds to the nonmagnetic and the twinned antiferromagnetic state, respectively. The representation of the twinned Fermi surface is based on a superposition of spectra obtained independently for antiferromagnetic states rotated by $90^\circ$ against each other which is  supposed to be a good approximation for twinned crystals, see for example the work of Tanatar \etal \cite{TKN+09}.

Comparing the nonmagnetic and the twinned antiferromagnetic state with each other it is obvious that there is a significant difference in the shape of the Fermi surface which is due to the underlying change of the electronic structure during the magnetic phase transition. However, the twinned Fermi surface is in principle isotropic along $\bv k_x$ and $\bv k_y$ due to the fact that the in-plane anisotropy cancels almost completely for two magnetic domains that are rotated by $90^\circ$ against each other. This means that although some influence of the magnetic ordering can be seen for twinned crystals, it is not possible to deduce information about the important in-plane anisotropy from the corresponding spectra. This stresses again the importance of ARPES measurements and calculations on detwinned crystals to investigate the magnetic structure correctly. To summarize, the agreement of the calculations with the experimental data is altogether quite well for the nonmagnetic as well as for the twinned magnetic state.

\begin{figure}[tb]
  \setlength{\unitlength}{0.61cm}\small
     \begin{picture}(5.9,8.8)
      \put(1.1,1.2){\includegraphics[height=6.8\unitlength,clip]{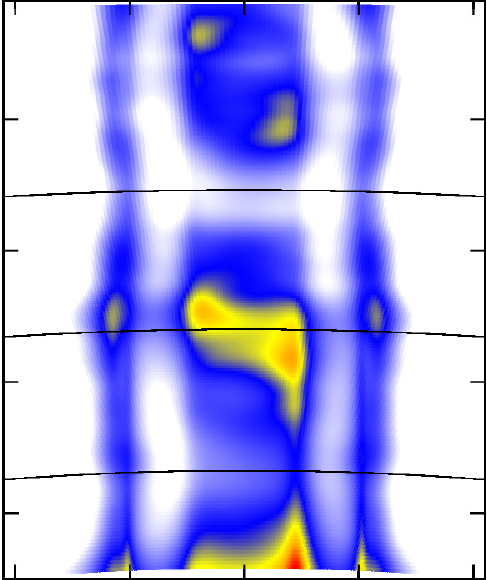}}     
      \multiputlist(0.4,1.95)(0,1.55)[l]{2.5,3.0,3.5,4.0}
      \multiputlist(0.88,0.9)(1.28,0)[l]{-0.4,-0.2,~0.0,~0.2,~0.4}
      \put(-0.5,3.7){\rotatebox{90}{$\bv k_{z} \left[ \text{\AA}^{-1} \right]$}}
      \put(2.8,-0.1){{$\bv k_{x} \left[ \text{\AA}^{-1} \right]$}}
      \put(0.65,8.3){\large (a)} \put(2.85,8.2){\large $a$-axis} 
      \footnotesize{\textcolor{black}{
      \put(6.1,5.8){48}\put(6.1,5.3){eV}\put(1.3,5.8){$ \Gamma$}
      \put(6.1,4.2){34}\put(6.1,3.7){eV}\put(1.3,4.2){${\mathrm Z}$}
      \put(6.1,2.5){22}\put(6.1,2.0){eV}\put(1.3,2.5){$ \Gamma$}}}
      \end{picture}
     \begin{picture}(7.0,8.8)
      \put(1.1,1.2){\includegraphics[height=6.8\unitlength,clip]{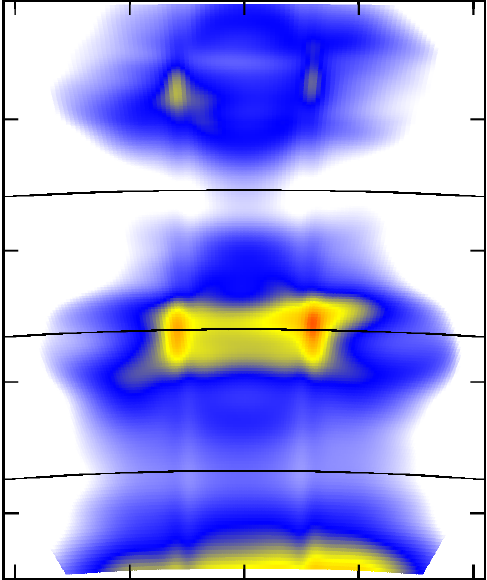}}
      \multiputlist(0.88,0.9)(1.28,0)[l]{-0.4,-0.2,~0.0,~0.2,~0.4}
      \put(2.8,-0.1){{$\bv k_y \left[ \text{\AA}^{-1} \right]$}}
      \put(0.65,8.3){\large (b)} \put(2.85,8.2){\large $b$-axis}     
      \footnotesize{\textcolor{black}{ 
      \put(1.3,5.8){$ \Gamma$}
      \put(1.3,4.2){${\mathrm Z}$}
      \put(1.3,2.5){$ \Gamma$}}}
      \scriptsize
      \put(6.9,1.2){\includegraphics[height=6.8\unitlength,clip]{figScale.pdf}}
      \put(7.25,8.2){\makebox(0,0){max}}
      \put(7.25,1.0){\makebox(0,0){min}}
      \end{picture}
\caption{(Color online) Calculated $\bv k_z$ dispersion as seen by ARPES along the both in-plane real space axes $a$ (a) and $b$ (b). The black lines mark the photon energies where the alternation of $\Gamma$ and ${\mathrm Z}$ can be seen along $\bv k_z$. Notably, the vertical intensity stripes at $\bv k_x \approx \pm 0.2$\;\AA ~in (a) seem almost independent on $\bv k_z$, indicating some connection to a surface related phenomenon.
}\label{Fig_Cal_KZ}
\end{figure}

\begin{figure}[b]
  \setlength{\unitlength}{0.61cm}\small
     \begin{picture}(5.9,8)
      \put(1.10,1.2){\includegraphics[height=5.5\unitlength,clip]{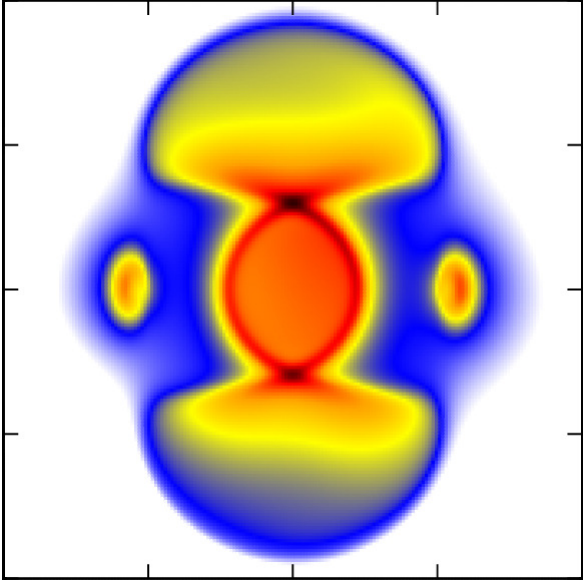}}
      \multiputlist(0.2,1.35)(0,1.3)[l]{-0.4,-0.2,~0.0,~0.2,~0.4}
      \multiputlist(0.75,0.9)(1.3,0)[l]{-0.4,-0.2,~0.0,~0.2,~0.4}
      \put(-0.6,3.0){\rotatebox{90}{$\bv k_y \left[ \text{\AA}^{-1} \right]$}}
      \put(2.7,-0.1){{$\bv k_x \left[ \text{\AA}^{-1} \right]$}}
      \put(0.65,7.0){\large  (a)}
      \put(2.45,7.0){\large $h\nu=34$\;eV}
      \end{picture}
     \begin{picture}(7.3,8)
      \put(1.20,1.2){\includegraphics[height=5.5\unitlength,clip]{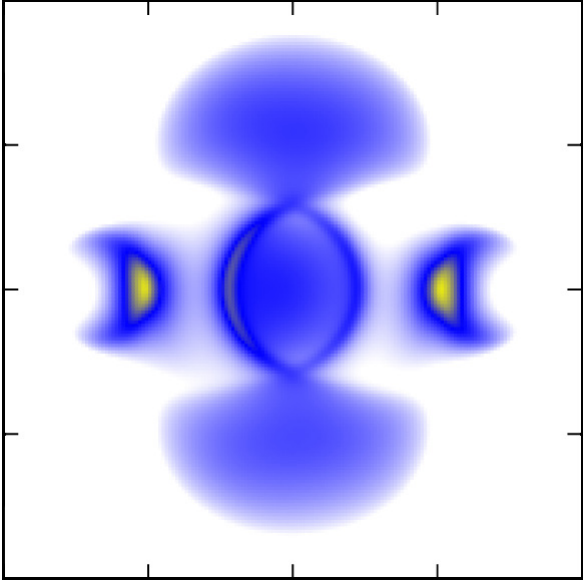}}
      \multiputlist(0.85,0.9)(1.3,0)[l]{-0.4,-0.2,~0.0,~0.2,~0.4}
      \put(2.8,-0.1){{$\bv k_x \left[ \text{\AA}^{-1} \right]$}}
      \put(0.65,6.95){\large (b)}
      \put(2.45,6.95){\large $h\nu=48$\;eV}
      \scriptsize
      \put(7.0,1.2){\includegraphics[height=5.5\unitlength,clip]{figScale.pdf}}
      \put(7.3,6.95){\makebox(0,0){max}}
      \put(7.3,1.0){\makebox(0,0){min}}
      \end{picture}
\caption{(Color online) Calculated Fermi surfaces for two additional photon energies $h\nu=34$\;eV and $h\nu=48$\;eV , corresponding to either ${\mathrm Z}$ (a) or to $\Gamma$ (b). It can be seen that the topology of interest, namely the anisotropic features are principally independent on $\bv k_z$.}\label{Fig_Cal_3448}
\end{figure}

Going back to the original study of detwinned antiferromagnetic crystals the $\bv k_z$ dispersion is shown along the $a$- and $b$-axes in Fig.~\ref{Fig_Cal_KZ}. The difference between $\Gamma$ and ${\mathrm Z}$ manifests itself mainly by the alternating intensity distributions. We find $\Gamma$ for photon energies of $h\nu=$22\;eV and $h\nu=48$\;eV respectively, while ${\mathrm Z}$ can be found at $h\nu=34\;eV$. This is in good agreement with literature which reports ${\mathrm Z}$ at $h\nu=35$\;eV and $\Gamma$ at $h\nu=49$\;eV.\cite{KFK+10} The anisotropic features, namely the bright spots along $\bv k_x$ and the pedals along $\bv k_y$ seem quite independent on $\bv k_z$, which agrees with the experimental reports on the detwinned crystals.\cite{YLC+11} For further discussion the Fermi surfaces for $h\nu=34$\;eV and $h\nu=48$\;eV respectively are shown in Fig.~\ref{Fig_Cal_3448}. The important aspect to note is that the anisotropic features are preserved independent on $\bv k_z$, meaning they are preserved for $\Gamma$ as well as for ${\mathrm Z}$. However, the most striking anisotropy between the $a$- and $b$-directions seen in the $\bv k_z$ dispersion are the almost vertical intensity lines along the $a$-axis for $\bv k_x \approx \pm 0.2$\;\AA. They are surprisingly robust concerning the $\bv k_z$ dispersion, already indicating possible surface related phenomena, as will be discussed later in more detail.

%
\begin{figure*}[tb]
  \setlength{\unitlength}{0.78cm}\normalsize
     \begin{picture}(6.8,9.5)
      \put(1.1,1.2){\includegraphics[height=6.8\unitlength,clip]{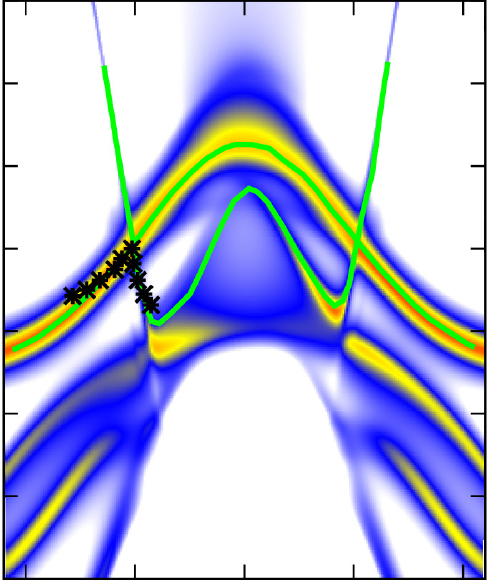}}     
      \multiputlist(0.2,1.25)(0,0.96)[l]{-0.4,-0.3,-0.2,-0.1,~0.0,~0.1,~0.2,~0.3}
      \multiputlist(1.05,0.9)(1.25,0)[l]{-0.4,-0.2,~0.0,~0.2,~0.4}
      \put(-0.4,3.5){\rotatebox{90}{Energy [eV]}}
      \put(-0.50,8.9){\Large \boldmath $\bv q_{||\textbf{FM}}$ \unboldmath}
      \put(1.55,8.3){\large (a)} \put(3.25,8.3){\large $a$-axis} 
      \end{picture}
      \begin{picture}(6.8,8.8)
      \put(1.1,1.2){\includegraphics[height=6.8\unitlength,clip]{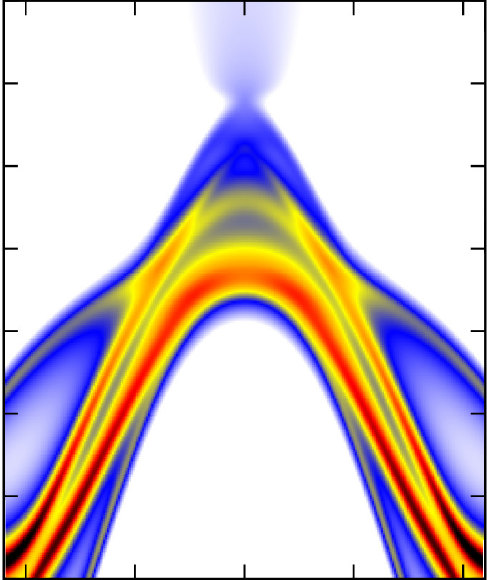}}
      \multiputlist(0.12,1.25)(0,0.96)[l]{-0.4,-0.3,-0.2,-0.1,~0.0,~0.1,~0.2,~0.3}
      \multiputlist(1.05,0.9)(1.25,0)[l]{-0.4,-0.2,~0.0,~0.2,~0.4}
      \put(1.55,8.3){\large (b)} \put(2.7,8.3){\large nonmagnetic}     
      \end{picture}
      \begin{picture}(7.8,8.8)
      \put(1.1,1.2){\includegraphics[height=6.8\unitlength,clip]{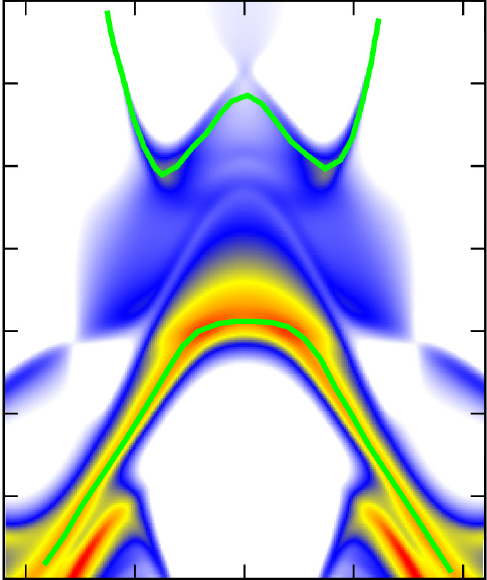}}
      \multiputlist(0.12,1.25)(0,0.96)[l]{-0.4,-0.3,-0.2,-0.1,~0.0,~0.1,~0.2,~0.3}
      \put(1.55,8.3){\large (c)} \put(3.25,8.3){\large $b$-axis}     
      \scriptsize
      \put(7.3,1.2){\includegraphics[height=6.8\unitlength,clip]{figScale.pdf}}
      \put(7.6,8.2){\makebox(0,0){max}}
      \put(7.6,1.0){\makebox(0,0){min}}
      \end{picture}
\\[-0.0cm]
     \begin{picture}(6.8,9.1)
      \put(1.1,1.2){\includegraphics[height=6.8\unitlength,clip]{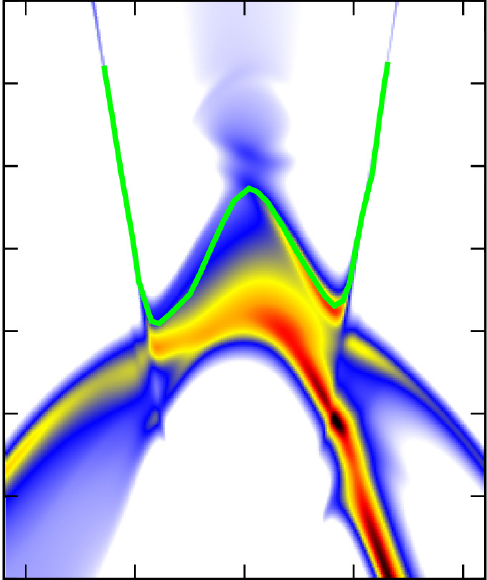}}     
      \multiputlist(0.2,1.25)(0,0.96)[l]{-0.4,-0.3,-0.2,-0.1,~0.0,~0.1,~0.2,~0.3}
      \multiputlist(1.05,0.9)(1.25,0)[l]{-0.4,-0.2,~0.0,~0.2,~0.4}
      \put(-0.4,3.5){\rotatebox{90}{Energy [eV]}}
      \put(3.2,-0.1){{$\bv k_{x} \left[ \text{\AA}^{-1} \right]$}}
      \put(-0.50,8.9){\Large \boldmath $\bv q_{||\textbf{AFM}}$ \unboldmath}
      \put(1.55,8.3){\large (d)} \put(3.25,8.3){\large $a$-axis} 
      \end{picture}
      \begin{picture}(6.8,8.8)
      \put(1.1,1.2){\includegraphics[height=6.8\unitlength,clip]{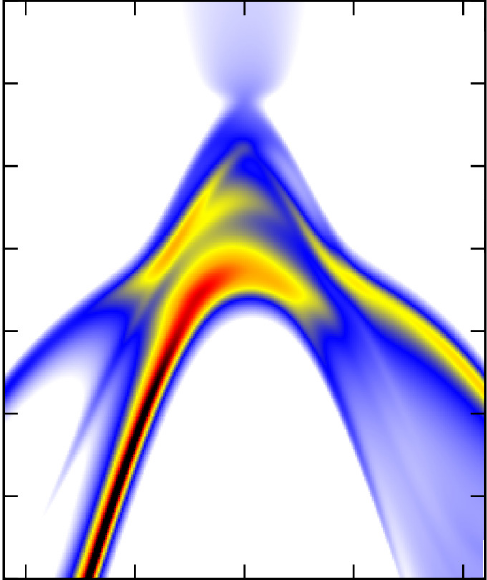}}
      \multiputlist(0.12,1.25)(0,0.96)[l]{-0.4,-0.3,-0.2,-0.1,~0.0,~0.1,~0.2,~0.3}
      \multiputlist(1.05,0.9)(1.25,0)[l]{-0.4,-0.2,~0.0,~0.2,~0.4}
      \put(3.2,-0.1){{$\bv k_y \left[ \text{\AA}^{-1} \right]$}}
      \put(1.55,8.3){\large (e)} \put(2.7,8.3){\large nonmagnetic}     
      \end{picture}
      \begin{picture}(7.8,8.8)
      \put(1.1,1.2){\includegraphics[height=6.8\unitlength,clip]{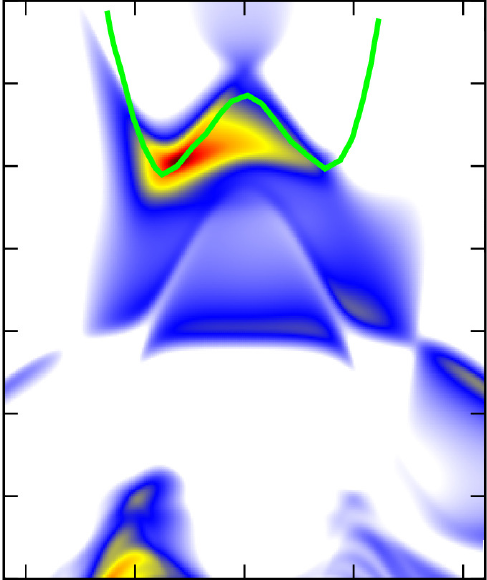}}
      \multiputlist(0.12,1.25)(0,0.96)[l]{-0.4,-0.3,-0.2,-0.1,~0.0,~0.1,~0.2,~0.3}
      \multiputlist(1.05,0.9)(1.25,0)[l]{-0.4,-0.2,~0.0,~0.2,~0.4}
      \put(3.2,-0.1){{$\bv k_y \left[ \text{\AA}^{-1} \right]$}}
      \put(1.55,8.3){\large (f)} \put(3.25,8.3){\large $b$-axis}     
      \scriptsize
      \put(7.3,1.2){\includegraphics[height=6.8\unitlength,clip]{figScale.pdf}}
      \put(7.6,8.2){\makebox(0,0){max}}
      \put(7.6,1.0){\makebox(0,0){min}}
      \end{picture}
\caption{(Color online) Calculated bands as seen by ARPES depending on the polarization of light, $\bv q_{||\text{FM}}$ (a-c) and $\bv q_{||\text{AFM}}$ (d-f), as well as on the orientation of the magnetic phase, either along the antiferromagnetic $a$-axis (a) and (d) or along the ferromagnetic $b$-axis (c) and (d). The band structure of the nonmagnetic phase is shown for comparison in (b) and (e). 
The solid green lines are guides to the eye for the important anisotropic bands in the magnetic phase. The black points in (a) are reproduced from experimental data of Yi \etal\cite{YLC+11} indicating the cut at the Fermi level for the two important anisotropic bands along the $a$-axis as seen by ARPES.}\label{Fig_Cal_BS00}
\end{figure*}

To complete the study of the in-plane anisotropy in the undoped compound the spin-dependent bands are investigated polarization-dependent along the two in-plane directions $a$ and $b$ for $h\nu=22$\;eV and for comparison the isotropic bands of the nonmagnetic case. Anisotropies due to the orthorhombic lattice distortion for the nonmagnetic case are very small and have no significant influence, as shown also in earlier work.\cite{DPM+14} 

The nonmagnetic bands for the polarizations $\bv q_{||\text{FM}}$ and $\bv q_{||\text{AFM}}$ are shown in Fig.~\ref{Fig_Cal_BS00} (b) and (e), respectively. 
Already for the nonmagnetic case it becomes obvious, that more information can be deduced for light with a polarization parallel to the ferromagnetic chains. For a perpendicular light polarization the intensity for some bands decrease so strongly that they practically seem to vanish. This is however not due to a vanishing of the bands but only due to the strong intensity variation, i.e. matrix element effects, as already mentioned before and as seen in experiment.

For the spin-polarized band structure with antiferromagnetic order along the $a$-axis the corresponding cases for $\bv q_{||\text{FM}}$ and $\bv q_{||\text{AFM}}$ are shown in Fig.~\ref{Fig_Cal_BS00} (a) and (d), respectively. The green solid lines are guides to the eye which emphasize the two important anisotropic bands in these spectra. First of all, there is some significant reorientation of the bands compared to the nonmagnetic case. Most striking is the appearance of a steep double-u shaped band which was not visible in the nonmagnetic case. This new appearance is most likely due to down-folding of the Brillouin zone when going from the 2-Fe cell to a magnetic 4-Fe cell. The second important band is a pure hole pocket which is compared to the nonmagnetic case shifted to higher binding energies. It should be noted that the intensity of this band is extremely polarization dependent. It is the dominating band for $\bv q_{||\text{FM}}$ while it is barely visible for $\bv q_{||\text{FM}}$. Comparing with the polarization dependent Fermi surface in Fig.~\ref{Fig_Cal_AFpol} it is obvious that this band is also part of the bright intensity spots in the Fermi surface along the $a$-direction and very characteristic for the anisotropy. It is also noteworthy that these two significant bands cross each other exactly at the Fermi level. This crossing is also reported in experiment as can be seen from the black points in Fig.~\ref{Fig_Cal_BS00} (a), which are reproduced from the experimental ARPES data of Yi~\etal \cite{YLC+11}. Thus, the experimental ARPES data could be again well reproduced by the calculations.

The situation for the magnetic bands with ferromagnetic order along the $b$-axis as shown in Fig.~\ref{Fig_Cal_BS00} (c) and (f) is in many aspects similar to that for the bands along the $a$-direction. One can identify two prominent anisotropic bands, one with a double-u like shape which has a higher intensity for a light polarization of $\bv q_{||\text{AFM}}$, while the other band marked with the solid green line is the dominating one for light $\bv q_{||\text{FM}}$. The important difference is that these bands do not touch each other as they are significantly shifted away in binding energy. Note that also no crossing is reported in experiment.\cite{YLC+11} Why these steep bands with the double-u shape cannot be seen in experiment for the $b$-direction gets also clear: The responsible band is simply completely shifted above the Fermi level. This observation can in principle be compared to the band splitting in ferromagnets. Note that along the $b$-axis there is ferromagnetic coupling while along the $a$-axis the magnetic order is antiferromagnetic. Thus, for \ce{BaFe2As2} one sees along the ferromagnetic chains a band splitting of approximately $0.2$\;eV for a magnetic moment around $0.7\muB$. This is comparable for example to Ni which shows a band splitting of approximately $0.3$\;eV for a moment of approximately $0.6\muB$.\cite{EHK80} Consequently, for decreasing magnetic moments upon alloying one expects a reduced band splitting together with a continuous matching of the anisotropic bands. To investigate this issue in further detail one has to look at the evolution of the ARPES band structure for increasing Co substitution on the Fe position which goes in hand with the reduction of the magnetic moments.

\subsection{Influence of Co substitution on the anisotropy}

\begin{figure*}[t]
  \setlength{\unitlength}{0.64cm}\small
     \begin{picture}(6.6,9.3)
      \put(1.1,1.2){\includegraphics[height=6.8\unitlength,clip]{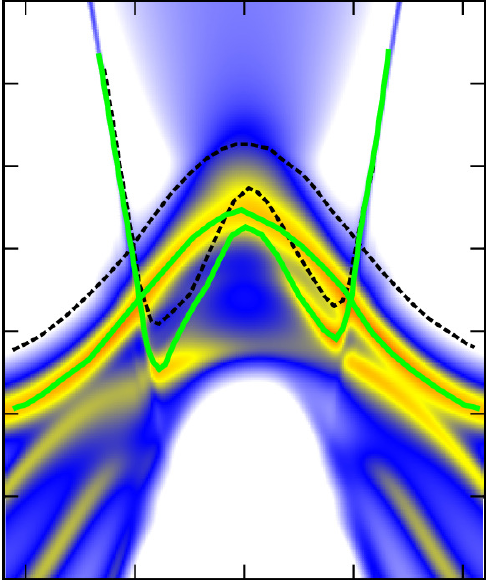}}     
      \multiputlist(0.2,1.25)(0,0.96)[l]{-0.4,-0.3,-0.2,-0.1,~0.0,~0.1,~0.2,~0.3}
      \multiputlist(1.05,0.9)(1.25,0)[l]{-0.4,-0.2,~0.0,~0.2,~0.4}
      \put(-0.4,3.5){\rotatebox{90}{Energy [eV]}}
      \put(3.2,-0.1){{$\bv k_{x} \left[ \text{\AA}^{-1} \right]$}}
      \put(-0.50,8.9){\Large \textbf{\textit{a}-axis}}
      \put(1.55,8.3){\large (a)} \put(3.0,8.3){\large $x=0.05$} 
      \end{picture}
     \begin{picture}(6.6,8.8)
      \put(1.1,1.2){\includegraphics[height=6.8\unitlength,clip]{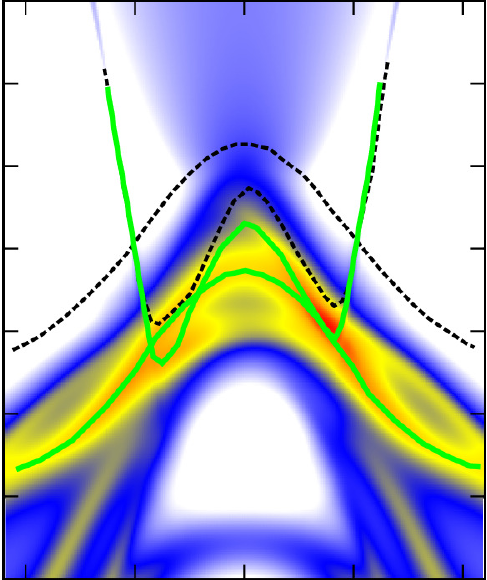}}
      \multiputlist(0.25,1.25)(0,0.96)[l]{-0.4,-0.3,-0.2,-0.1,~0.0,~0.1,~0.2,~0.3}
      \multiputlist(1.05,0.9)(1.25,0)[l]{-0.4,-0.2,~0.0,~0.2,~0.4}
      \put(3.2,-0.1){{$\bv k_x \left[ \text{\AA}^{-1} \right]$}}
      \put(1.55,8.3){\large (b)} \put(3.0,8.3){\large $x=0.10$}     
      \end{picture}
     \begin{picture}(6.6,8.8)
      \put(1.1,1.2){\includegraphics[height=6.8\unitlength,clip]{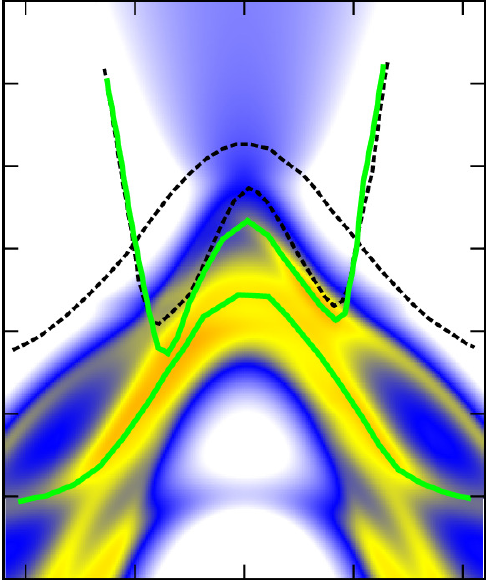}}
      \multiputlist(0.25,1.25)(0,0.96)[l]{-0.4,-0.3,-0.2,-0.1,~0.0,~0.1,~0.2,~0.3}
      \multiputlist(1.05,0.9)(1.25,0)[l]{-0.4,-0.2,~0.0,~0.2,~0.4}
      \put(3.2,-0.1){{$\bv k_x \left[ \text{\AA}^{-1} \right]$}}
      \put(1.55,8.3){\large (c)} \put(3.0,8.3){\large $x=0.125$}     
      \end{picture}
     \begin{picture}(7.0,8.8)
      \put(1.1,1.2){\includegraphics[height=6.8\unitlength,clip]{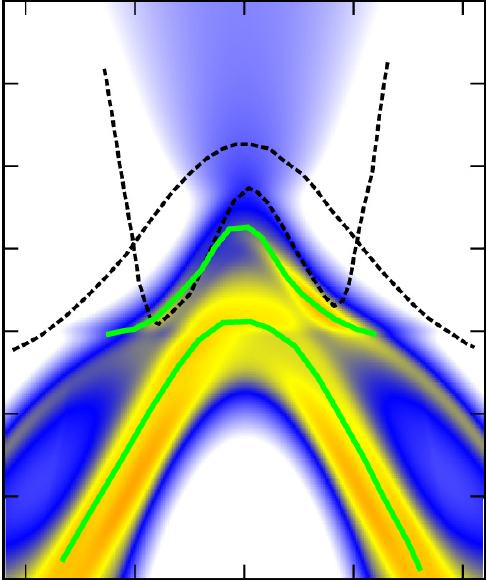}}
      \multiputlist(0.25,1.25)(0,0.96)[l]{-0.4,-0.3,-0.2,-0.1,~0.0,~0.1,~0.2,~0.3}
      \multiputlist(1.05,0.9)(1.25,0)[l]{-0.4,-0.2,~0.0,~0.2,~0.4}
      \put(3.2,-0.1){{$\bv k_x \left[ \text{\AA}^{-1} \right]$}}
      \put(1.55,8.3){\large (d)} \put(3.0,8.3){\large $x=0.15$}     
      \scriptsize
      \put(7.0,1.2){\includegraphics[height=6.8\unitlength,clip]{figScale.pdf}}
      \put(7.3,8.2){\makebox(0,0){max}}
      \put(7.3,1.0){\makebox(0,0){min}}
      \end{picture}
\\[+0.7cm]
     \begin{picture}(6.6,9.3)
      \put(1.1,1.2){\includegraphics[height=6.8\unitlength,clip]{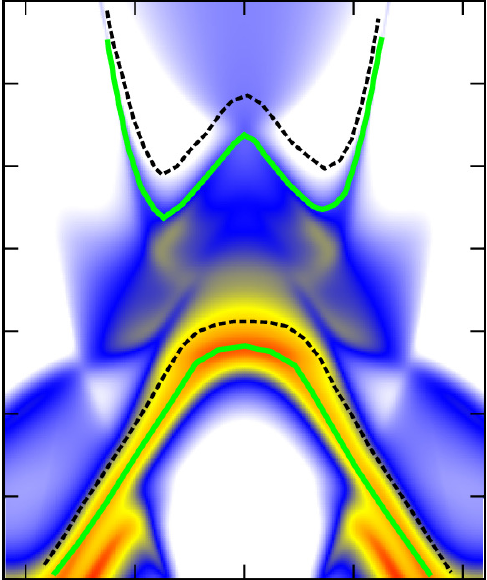}}     
      \multiputlist(0.2,1.25)(0,0.96)[l]{-0.4,-0.3,-0.2,-0.1,~0.0,~0.1,~0.2,~0.3}
      \multiputlist(1.05,0.9)(1.25,0)[l]{-0.4,-0.2,~0.0,~0.2,~0.4}
      \put(-0.4,3.5){\rotatebox{90}{Energy [eV]}}
      \put(3.2,-0.1){{$\bv k_y \left[ \text{\AA}^{-1} \right]$}}
      \put(-0.50,8.9){\Large \textbf{\textit{b}-axis}}
      \put(1.55,8.3){\large (e)} \put(3.0,8.3){\large $x=0.05$} 
      \end{picture}
     \begin{picture}(6.6,8.8)
      \put(1.1,1.2){\includegraphics[height=6.8\unitlength,clip]{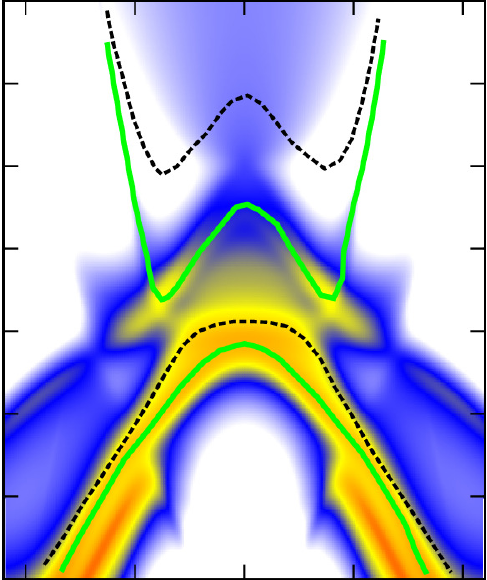}}
      \multiputlist(0.25,1.25)(0,0.96)[l]{-0.4,-0.3,-0.2,-0.1,~0.0,~0.1,~0.2,~0.3}
      \multiputlist(1.05,0.9)(1.25,0)[l]{-0.4,-0.2,~0.0,~0.2,~0.4}
      \put(3.2,-0.1){{$\bv k_y \left[ \text{\AA}^{-1} \right]$}}
      \put(1.55,8.3){\large (f)} \put(3.0,8.3){\large $x=0.10$}     
      \end{picture}
     \begin{picture}(6.6,8.8)
      \put(1.1,1.2){\includegraphics[height=6.8\unitlength,clip]{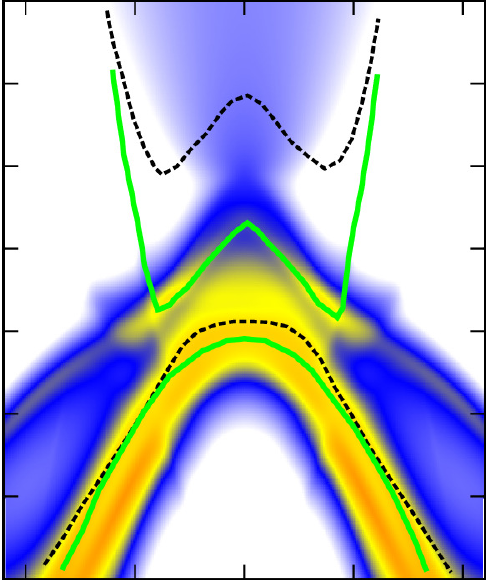}}
      \multiputlist(0.25,1.25)(0,0.96)[l]{-0.4,-0.3,-0.2,-0.1,~0.0,~0.1,~0.2,~0.3}
      \multiputlist(1.05,0.9)(1.25,0)[l]{-0.4,-0.2,~0.0,~0.2,~0.4}
      \put(3.2,-0.1){{$\bv k_y \left[ \text{\AA}^{-1} \right]$}}
      \put(1.55,8.3){\large (g)} \put(3.0,8.3){\large $x=0.125$}     
      \end{picture}
     \begin{picture}(7.0,8.8)
      \put(1.1,1.2){\includegraphics[height=6.8\unitlength,clip]{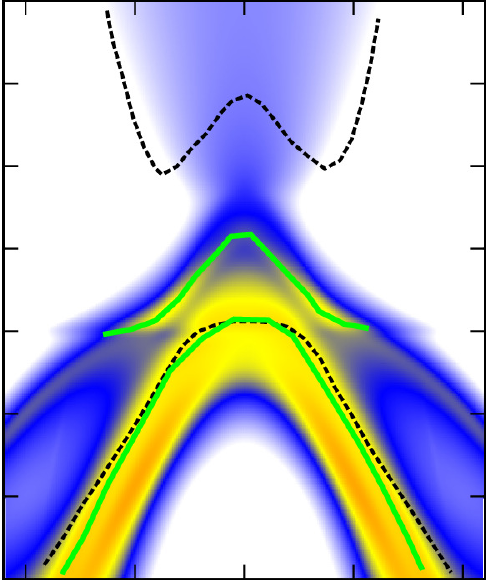}}
      \multiputlist(0.25,1.25)(0,0.96)[l]{-0.4,-0.3,-0.2,-0.1,~0.0,~0.1,~0.2,~0.3}
      \multiputlist(1.05,0.9)(1.25,0)[l]{-0.4,-0.2,~0.0,~0.2,~0.4}
      \put(3.2,-0.1){{$\bv k_y \left[ \text{\AA}^{-1} \right]$}}
      \put(1.55,8.3){\large (h)} \put(3.0,8.3){\large $x=0.15$}     
      \scriptsize
      \put(7.0,1.2){\includegraphics[height=6.8\unitlength,clip]{figScale.pdf}}
      \put(7.3,8.2){\makebox(0,0){max}}
      \put(7.3,1.0){\makebox(0,0){min}}
      \end{picture}
\caption{(Color online) Calculated bands as seen by ARPES for different Co concentrations $x$ in magnetic \ce{Ba(Fe_{1-$x$}Co_{$x$})2As2} depending on the axes $a$ (a-d) and $b$ (e-h) with a constant light polarization of $\bv q_{||\text{FM}}$ to easily identify the important bands. The black dashed lines correspond to the initial position of the anisotropic bands in the undoped compound, see Fig.~\ref{Fig_Cal_BS00}. The green solid lines are guides to the eye for the corresponding important anisotropic bands to see their change under increasing Co substitution. Their anisotropy vanishes together with the long-range antiferromagnetic order at $x=0.15$.}\label{Fig_Cal_BSCo}
\end{figure*}

Substitution of Fe with Co in \ce{Ba(Fe_{1-$x$}Co_{$x$})2As2} is one of the common ways to induce superconductivity in \ce{BaFe2As2} by electron doping. The substitution does consequently diminish the strength of the antiferromagnetic coupling within this compound until the long-range magnetic order collapses and superconductivity emerges.\cite{SJM+08,CAK+09} As the strength of the magnetic order decreases with Co doping, in experiment as well as in the calculations, it can be assumed that the strong in-plane anisotropy does also decrease. The breakdown of the long-range antiferromagnetic order in Fig.~\ref{Fig_MagnMom} appears for a somewhat higher Co concentrations than in experiment. 
Thus, the breakdown of the anisotropy is expected at higher doping levels. This is true for a concentration of $x=0.15$ in \ce{Ba(Fe_{1-$x$}Co_{$x$})2As2} which is the first substitution level where the magnetic order did completely vanish in the self-consistent calculation. Considering the evolution of the magnetic moments in Fig.~\ref{Fig_MagnMom} one can see that the initial magnetic moment decreased to approximately $75\%$, $50\%$ and $25\%$ of the original value for Co substitutions of $x=0.05$, $0.10$ and $0.125$ respectively. 

To investigate the impact of alloying in detail, the ARPES band structure for these concentrations including the already nonmagnetic $x=0.15$ for both directions $a$ and $b$ is presented in Fig.~\ref{Fig_Cal_BSCo}. The calculations are performed for $h\nu=22$\;eV comparable to Fig.~\ref{Fig_Cal_BS00} but they are only shown for a light polarization of $\bv q_{||\text{FM}}$ because is was already clarified that the anisotropic bands can be best seen with this specific polarization. The black dashed lines in Fig.~\ref{Fig_Cal_BSCo} are shown for comparison and they correspond to the band position of the anisotropic bands in the undoped compound \ce{BaFe2As2} with the highest anisotropy, seen in Fig.~\ref{Fig_Cal_BS00} (a) and (c). The green solid lines are guides to the eye to identify more easily the corresponding anisotropic bands for the specific Co concentrations. The difference between black dashed lines and green solid lines is thus the change of the original anisotropy with increasing Co substitution. For the case of the nonmagnetic \ce{Ba(Fe_{1-$x$}Co_{$x$})2As2} with $x=0.15$ shown in Fig.~\ref{Fig_Cal_BSCo} (d) and (h) the anisotropy has completely vanished and the band structures coincide with each other. This could be expected from experiment and it is reproduced in the calculations. It should be noted again at this point, that the crystal lattice is still orthorhombic, however, the lattice anisotropy is indeed too weak to be visible in the band structure.\cite{DPM+14}

Some other interesting findings can be deduced from the evolution of the band structure upon Co substitution. First of all the intensity of the double-u shaped band decreases continuously, however, it only completely disappears after the collapse of the long-range antiferromagnetic order. The change in anisotropy for the antiferromagnetic order along the $a$-axis is mostly characterized with the consequent shift of the hole-pocket to lower binding energies. This is also experimentally reported for a decrease in the magnetic coupling strength, either induced through Co doping or increasing temperature.\cite{YLC+11} Concerning this situation for the ferromagnetic order along the $b$-axis the most prominent feature is the shift of the double-u shaped band to lower binding energies. What can be seen in Fig.~\ref{Fig_Cal_BS00} (e) to (h) is that the energy difference of these two main anisotropic bands does strongly and continuously decrease. This is in agreement with the assumption of a smaller band splitting for decreasing ferromagnetic coupling strength.

In summary one can say that for the antiferromagnetic order along the $a$-axis mostly the hole-pocket changes while the double-u shaped band stays more or less constant. For the ferromagnetic order along the $b$-axis it is the other way round. The double-u shaped band undergoes the strongest change while the other band stays more or less unchanged in energy and shape. The final result is the same in both cases, a matching of the bands and a consequent isotropic in-plane band structure. This detailed analysis allows to follow the change from the strong in-plane anisotropy of the undoped compound to the isotropic behavior in the Co substituted system in a continuous way with direct correspondence to ARPES. Thus, this approach based on KKR-CPA proves its advantages for investigating the iron pnictide superconductors at regions of interest which are difficult to evaluate by means of other band structure methods. 

\subsection{Surface termination}

\begin{figure*}[tb]
  \setlength{\unitlength}{0.61cm}\small
\begin{tabular}{c|c}
      {\begin{picture}(6.2,9.3)
      \put(1.1,1.2){\includegraphics[height=5.8\unitlength,clip]{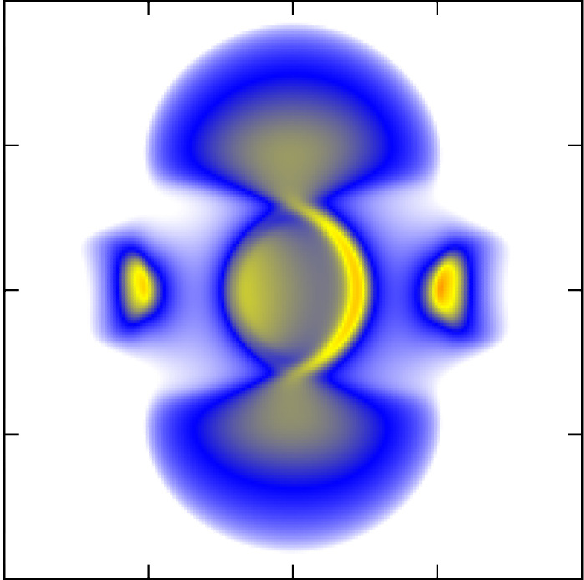}}     
      \multiputlist(0.2,1.15)(0,1.48)[l]{-0.4,-0.2,~0.0,~0.2,~0.4}
      \multiputlist(2.10,0.9)(1.45,0)[l]{-0.2,~0.0,~0.2}
      \put(-0.5,2.8){\rotatebox{90}{$\bv k_y \left[ \text{\AA}^{-1} \right]$}}
      \put(3.0,-0.0){{$\bv k_{x} \left[ \text{\AA}^{-1} \right]$}}
      \put(4.50,8.6){\Large \textbf{As-terminated}}
      \put(2.5,7.5){\large (a)} \put(5.5,7.5){\large Fermi surface} 
      \put(7.0,1.2){\includegraphics[height=5.8\unitlength,clip]{figScale.pdf}}
      \scriptsize
      \put(7.1,7.2){\makebox(0,0){max}}
      \put(7.1,1.0){\makebox(0,0){min}}
      \end{picture}
     \begin{picture}(7.5,8.8)
      \put(1.1,1.2){\includegraphics[height=5.8\unitlength,clip]{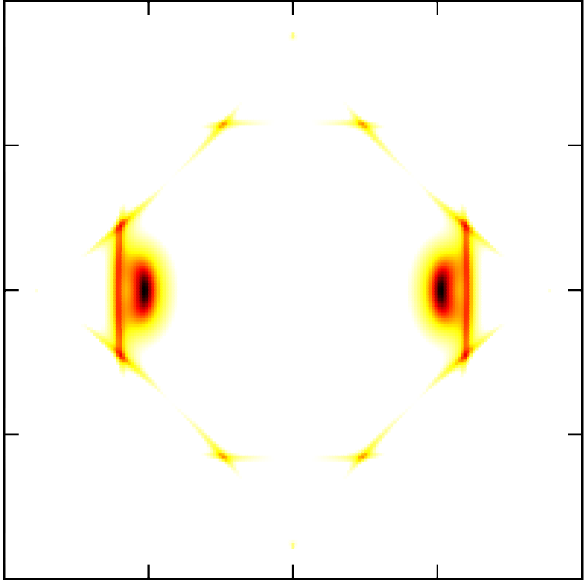}}
      \multiputlist(2.10,0.9)(1.45,0)[l]{-0.2,~0.0,~0.2}
      \put(3.0,-0.0){{$\bv k_x \left[ \text{\AA}^{-1} \right]$}}
      \put(7.0,1.2){\includegraphics[height=5.8\unitlength,clip]{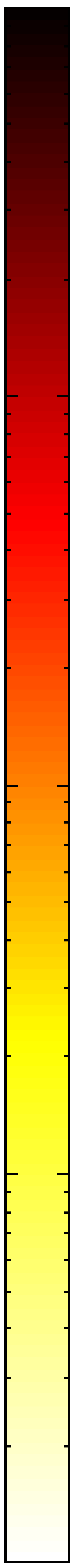}}   
      \scriptsize
      \put(7.1,7.2){\makebox(0,0){max}}
      \put(7.1,1.0){\makebox(0,0){min}}
      \end{picture}} 
\hspace{0.3cm} & \hspace{0.05cm}
     {\begin{picture}(6.2,9.3)
      \put(1.1,1.2){\includegraphics[height=5.8\unitlength,clip]{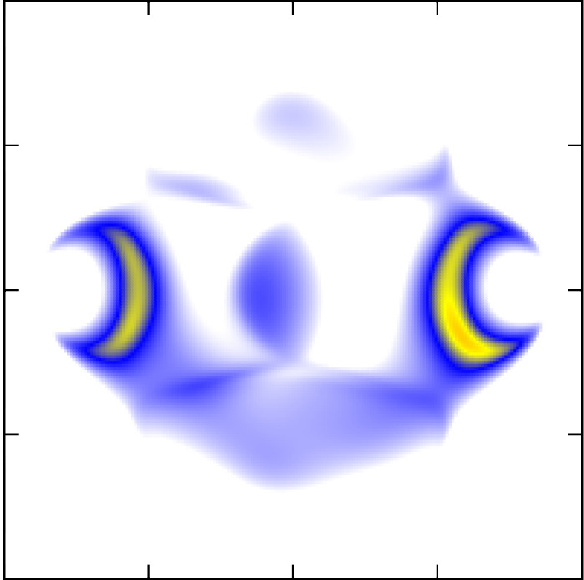}}     
      \multiputlist(0.2,1.15)(0,1.48)[l]{-0.4,-0.2,~0.0,~0.2,~0.4}
      \multiputlist(2.10,0.9)(1.45,0)[l]{-0.2,~0.0,~0.2}
      \put(3.0,-0.0){{$\bv k_{x} \left[ \text{\AA}^{-1} \right]$}}
      \put(4.50,8.6){\Large \textbf{Ba-terminated}}
      \put(2.5,7.5){\large (c)} \put(5.5,7.5){\large Fermi surface} 
      \put(7.0,1.2){\includegraphics[height=5.8\unitlength,clip]{figScale.pdf}}
      \scriptsize
      \put(7.1,7.2){\makebox(0,0){max}}
      \put(7.1,1.0){\makebox(0,0){min}}
      \end{picture}
     \begin{picture}(7.5,8.8)
      \put(1.1,1.2){\includegraphics[height=5.8\unitlength,clip]{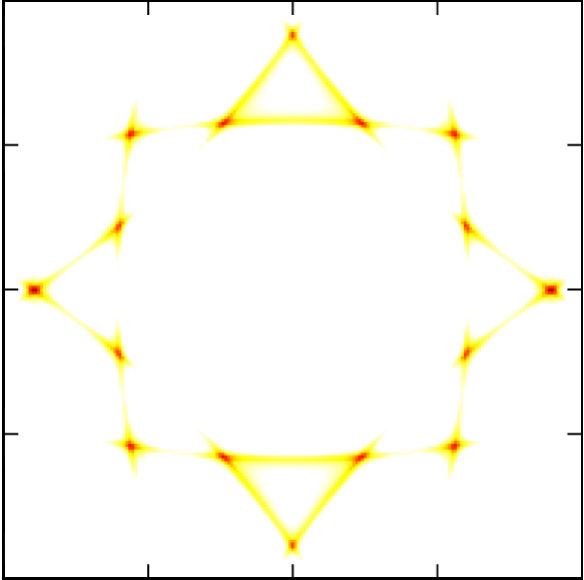}}
      \multiputlist(2.10,0.9)(1.45,0)[l]{-0.2,~0.0,~0.2}
      \put(3.0,-0.0){{$\bv k_x \left[ \text{\AA}^{-1} \right]$}}
      \put(7.0,1.2){\includegraphics[height=5.8\unitlength,clip]{figScale2.pdf}}   
      \scriptsize
      \put(7.1,7.2){\makebox(0,0){max}}
      \put(7.1,1.0){\makebox(0,0){min}}
      \end{picture}} 
\\[+0.4cm]
     {\begin{picture}(6.55,9.3)
      \put(1.1,1.2){\includegraphics[height=6.8\unitlength,clip]{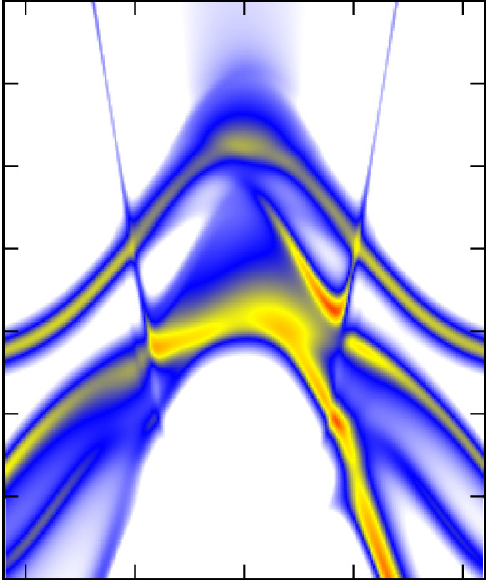}}     
      \multiputlist(0.2,1.25)(0,0.96)[l]{-0.4,-0.3,-0.2,-0.1,~0.0,~0.1,~0.2,~0.3}
      \multiputlist(0.9,0.9)(1.25,0)[l]{-0.4,-0.2,~0.0,~0.2,~0.4}
      \put(-0.4,3.5){\rotatebox{90}{Energy [eV]}}
      \put(3.0,-0.0){{$\bv k_{x} \left[ \text{\AA}^{-1} \right]$}}
      \put(2.0,8.5){\large (b)} \put(3.0,8.5){\large Band structure along $a$-axis} 
      \put(7.05,1.2){\includegraphics[height=6.8\unitlength,clip]{figScale.pdf}}
      \scriptsize
      \put(7.25,8.2){\makebox(0,0){max}}
      \put(7.25,1.1){\makebox(0,0){min}}
      \end{picture}
     \begin{picture}(7.4,8.8)
      \put(1.1,1.2){\includegraphics[height=6.8\unitlength,clip]{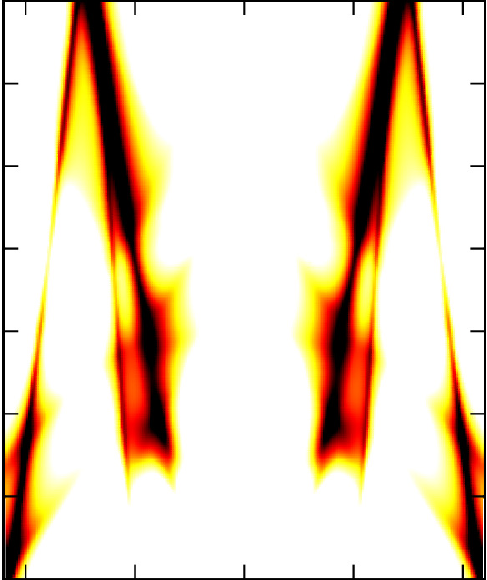}}
      \multiputlist(0.9,0.9)(1.25,0)[l]{-0.4,-0.2,~0.0,~0.2,~0.4}
      \put(3.0,-0.0){{$\bv k_x \left[ \text{\AA}^{-1} \right]$}}  
      \put(7.0,1.2){\includegraphics[height=6.8\unitlength,clip]{figScale2.pdf}}   
      \scriptsize
      \put(7.25,8.2){\makebox(0,0){max}}
      \put(7.22,1.0){\makebox(0,0){min}}
      \end{picture}} 
\hspace{0.15cm} & \hspace{-0.15cm}
     {\begin{picture}(6.55,9.3)
      \put(1.1,1.2){\includegraphics[height=6.8\unitlength,clip]{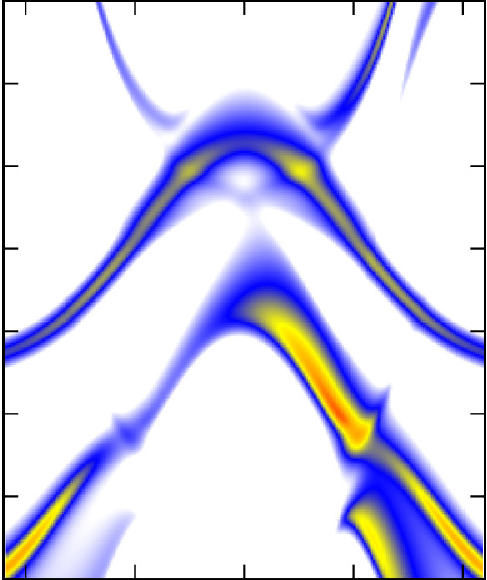}}     
      \multiputlist(0.2,1.25)(0,0.96)[l]{-0.4,-0.3,-0.2,-0.1,~0.0,~0.1,~0.2,~0.3}
      \multiputlist(0.9,0.9)(1.25,0)[l]{-0.4,-0.2,~0.0,~0.2,~0.4}
      \put(3.0,-0.0){{$\bv k_{x} \left[ \text{\AA}^{-1} \right]$}}
      \put(2.0,8.5){\large (d)} \put(3.0,8.5){\large Band structure along $a$-axis} 
      \put(7.05,1.2){\includegraphics[height=6.8\unitlength,clip]{figScale.pdf}}
      \scriptsize
      \put(7.25,8.2){\makebox(0,0){max}}
      \put(7.25,1.1){\makebox(0,0){min}}
      \end{picture}
     \begin{picture}(7.4,8.8)
      \put(1.1,1.2){\includegraphics[height=6.8\unitlength,clip]{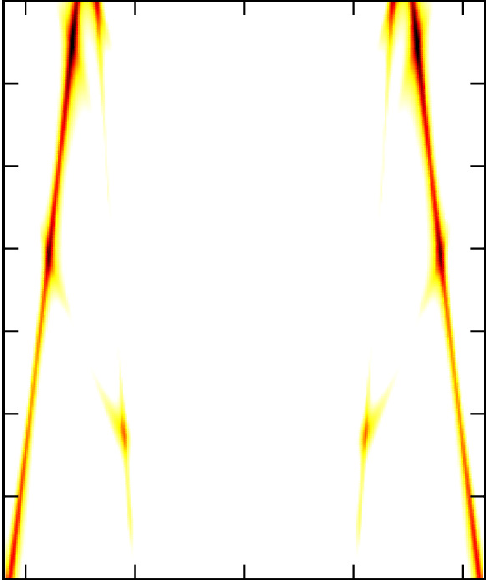}}
      \multiputlist(0.9,0.9)(1.25,0)[l]{-0.4,-0.2,~0.0,~0.2,~0.4}
      \put(3.0,-0.0){{$\bv k_x \left[ \text{\AA}^{-1} \right]$}}  
      \put(7.0,1.2){\includegraphics[height=6.8\unitlength,clip]{figScale2.pdf}}   
      \scriptsize
      \put(7.25,8.2){\makebox(0,0){max}}
      \put(7.22,1.0){\makebox(0,0){min}}
      \end{picture}} 
\end{tabular}
\caption{(Color online) Calculated Fermi surfaces and band structures for either an As-terminated surface (a) and (b) or a Ba-terminated surface (c) and (d). The right side of each picture shows the corresponding plot of $1/|D(E, \bv k)|$, meaning a high intensity indicates a possible surface state if this specific structure can be also identified in the regular electronic structure calculation. Clear surface states can be identified for the As-terminated surface as bright spots in der Fermi surface (a) and as corresponding steep bands in the band structure (b). These surface states are missing for a Ba-terminated surface.}\label{Fig_Cal_Det}
\end{figure*}

Using the one-step model of photoemission one can identify different surface states and can thus clarify the origin of surface bands. The reason for the occurrence of surface-states has long been developed in multiple scattering theory, which is the underlying basis of the SPR-KKR method.\cite{EP78,McR79} The so-called determinant condition uses the reflection matrices of the bulk crystal $\bv R_\text{b}$ and of the surface barrier potential $\bv R_\text{v}$, which connects the inner potential of the bulk crystal with the vacuum level. The appearance of a surface state is given by the following condition:
\begin{equation}
 D(E, \bv k) = \det \left( 1\!\!1  - \bv R_\text{b}(E, \bv k)\bv R_\text{v}(E, \bv k) \right) = 0. 
\end{equation}
For better visualization we plot $1/|D(E, \bv k)|$ in the following. If this expression is bigger than approximately $10^3$ we speak about a surface state. For values between $10^0$ and $10^3$ the state is defined as a so-called surface resonance. For values below one has bulk states. More details can be found in the overview by Braun and Donath.\cite{BD02}

The application of this determinant approach is demonstrated in Fig.~\ref{Fig_Cal_Det}. Here one can see the Fermi surfaces and the band structures along the $a$-axis for an As-terminated and a Ba-terminated surface, respectively, together with the corresponding plot of $1/|D(E, \bv k)|$ on the right hand side of each picture. The bands are shown for $h\nu=22$\;eV and they are averaged over the two light polarizations $\bv q_{||\text{FM}}$ and $\bv q_{||\text{AFM}}$ in order to make all relevant contributions equally visible. 

It should be noted, that the determinant condition itself and without a high intensity in the corresponding band structure plot is only an indication for a surface state or a surface resonance. Only if a high intensity in the $1/|D(E, \bv k)|$ plot coincides with a band in the band structure one can associate this band with a clear surface character. For example has the bright octagon shape of the determinant plot of the Fermi surface in Fig.~\ref{Fig_Cal_Det} (a) not a corresponding counterpart in the band structure plot. The two high intensity spots along the $a$-axis which are equally visible in the Fermi surface as well as in the determinant condition are in clear contrast to this behavior. Thus, this feature has a surface related origin, more specifically a surface state as the intensity of $1/|D(E, \bv k)|$ is in the order of $10^6$.

This surface state can be also identified in the band structure along the $a$-axis as shown in Fig.~\ref{Fig_Cal_Det} (b) where a strong intensity in the determinant plot coincides with the steep bands that cut the Fermi level and which are part of the already discussed double-u shape. Consequently, these bands can be identified as surface states. This is in accordance with the earlier findings for the $\bv k_z$-dispersion in Fig.~\ref{Fig_Cal_KZ} (a) where the vertical intensities at $\bv k_x \approx \pm 0.2$\;\AA ~were independent on $\bv k_z$, indicating a connection to a surface related phenomenon.

Another verification for the surface related origin of these bands is shown in Fig.~\ref{Fig_Cal_Det} (c) and (d), where the corresponding Fermi surface and band structure are shown for a Ba-terminated surface. As already indicated before all other calculations presented in this paper are under the assumption of an As-terminated surface. Obviously, the assumed surface termination has also an influence on surface related phenomena and surface states might be shifted significantly in energy. Indeed, the surface states discussed for the As-terminated surface have completely vanished in the Ba-terminated case. The corresponding high intensities are missing in the $1/|D(E, \bv k)|$ plots and in the band structure of Fig.~\ref{Fig_Cal_Det} (d) the characteristic steep bands from the As-terminated surface have also vanished. No intensities in the determinant plot coincide with features in the band structure and thus surface effects have been removed by the Ba termination. Overall, the Fermi surface and the band structure have undergone significant changes for the altered surface termination. The characteristic anisotropic features of the Fermi surface in Fig.~\ref{Fig_Cal_Det} (a) are hardly visible in the Ba-terminated case in Fig.~\ref{Fig_Cal_Det} (c). It seems like the Ba layer on top acts as some kind of damping layer which reduces the intensity and blurs the electronic states which are clearly visible in an As-terminated surface. In particular one has to note that the agreement with experimental ARPES data is significantly better for an As-terminated surface compared to the Ba-terminated one. Especially the steep bands along the $a$-axis are seen in experiment\cite{YLC+11} and they could be successfully identified as surface states which are only visible for an As-terminated surface.

This result can be used for conclusions on the most likely surface termination of \ce{BaFe2As2}. Interestingly, the surface termination in this material is still not clear and under debate, although several experimental measurements and theoretical calculations exist.\cite{Hof11} According to first principle calculations only three possibilities for the surface termination exist, namely a fully As-terminated or a fully Ba-terminated surface as well as an As surface covered with half of the stoichiometric Ba atoms.\cite{PFGC10} There are experimental scanning tunneling microscopy (STM) and low-energy electron-diffraction (LEED) measurements which indicate a Ba-terminated surface\cite{MJH+09}. However, there are also experimental STM + LEED data which clearly favor an As-terminated surface\cite{NLJ+09}. The ARPES calculations clearly favor an As-terminated surface as it was shown that agreement with experiment is considerably better compared to the Ba-terminated one. This cannot rule out the possibility of some partial covering with few Ba adatoms but one can state that every Ba atoms on top muffles the electronic structure seen in experiment and that this structure is due to an As-terminated compound. So one would expect a more or less clean As-terminated surface as most probable surface termination for \ce{BaFe2As2}. Additional covering with some Ba atoms might be possible but also might have some degrading influence on the quality of the ARPES measurement. 

%
\section{Summary}

The Munich SPR-KKR package was used for self-consistent and fully relativistic calculations of orthorhombic \ce{Ba(Fe_{1-$x$}Co_{$x$})2As2} in its experimentally observed stripe antiferromagnetic ground-state for $x=0.0$ up to $x=0.15$. The substitutional disorder induced by Co on Fe positions was dealt with on a CPA level which was earlier shown to be fully equal to a comprehensive supercell calculation.\cite{DPM+14} Magnetic moments of $0.73\muB$ for undoped \ce{BaFe2As2} were reproduced and additionally a reasonable magnetic behavior for increasing Co substitution with a continuous decrease of the magnetic moments until a collapse of the antiferromagnetic order at 15\% Co concentration was reached. This is in good agreement with experimental behavior.\cite{CAK+09,RTS+09,HQB+08}

Concerning ARPES most experimental data available is actually insufficient to talk about in-plane anisotropy due to twinning effects during the phase transition from the nonmagnetic tetragonal to the antiferromagnetic orthorhombic phase. A complicated detwinning process, typically with uniaxial stress on the single crystal, is necessary to gain anisotropic data of the electronic structure.\cite{YLC+11,KOK+11} Referring to the available experimental data it was possible to reproduce the electronic structure of \ce{BaFe2As2} in very good agreement with experiment. The Fermi surface shows all important anisotropic features, namely some bright spots of intensity along the  antiferromagnetic order along the $a$-axis and more blurred pedals along the ferromagnetic order along the $b$-axis. Also in agreement with experiment a strong dependence on the polarization of light was found, been either parallel to the ferromagnetic or to the antiferromagnetic order, indicating the strong multiorbital character. For comparison the Fermi surface of the nonmagnetic phase as well as a hypothetical Fermi surface for a twinned ARPES measurement as a superposition of two antiferromagnetic cells rotated by $90^\circ$ to each other was shown. Both were again in agreement with experiment. In addition to the anisotropic $\bv k_z$ dispersion some focus was put on the anisotropic band structure along the $a$- and $b$-axes and it was compared to the nonmagnetic case. One could identify the important anisotropic bands and these could be interpreted in terms of band splitting for the ferromagnetic chains along the $b$-axis, principally comparable to typical ferromagnetic band splitting as observed for example in Ni. 

In addition the evolution of these anisotropic bands for small steps of $x$ in \ce{Ba(Fe_{1-$x$}Co_{$x$})2As2} was presented until the breakdown of long-range antiferromagnetic order. The decreasing band splitting and a continuous matching of the anisotropic bands could be reproduced in great detail and consistent with experimental findings. 

Finally the so-called determinant condition $1/|D(E, \bv k)|$ was used to evaluate possible surface states of the band structure. It was possible to identify steep bands along the $a$-axis as surface states. These are at least partially responsible for the characteristic bright intensity spots in the electronic structure along the $a$-axis and can also be seen in experiment. Interestingly, these surface states are only visible near the Fermi level for an As-terminated surface. It was shown that a Ba-terminated top-layer acts as some kind of damping which moves the surface states far away and blurs the electronic structure. Significantly better agreement with experimental data is found for an As-terminated surface. This leads to the conclusion that an As-terminated surface would be most likely, an issue that is in fact experimentally not convincingly clarified.\cite{MJH+09,NLJ+09,Hof11} Some Ba adatoms might be still possible but one would expect a negative influence on the quality of the measurements. 

To conclude, this publication was successful in reproducing the strong in-plane anisotropy of \ce{BaFe2As2} and its behavior under substitution in very good agreement with experiment using ARPES calculations. These calculations allow even predictions on possible surface terminations. 

%
%
\section*{Acknowledgments}
We acknowledge the financial support from the  Deutsche Forschungsgemeinschaft DFG (projects FOR 1346) and from the Bundesministerium f\"ur Bildung und Forschung BMBF (project 05K13WMA). We further thank for the support from CENTEM PLUS (L01402).
%



\begin{thebibliography}{10}

\bibitem{KWHH08}
Y.~Kamihara, T.~Watanabe, M.~Hirano, and H.~Hosono,
\newblock J. Amer. Chem. Soc. {\bf 130}, 3296 (2008).

\bibitem{TIA+08}
H.~Takahashi, K.~Igawa, K.~Arii, Y.~Kamihara, M.~Hirano, and H.~Hosono,
\newblock Nature {\bf 453}, 376 (2008).

\bibitem{SLS+09}
Y.~Su, P.~Link, A.~Schneidewind, T.~Wolf, P.~Adelmann, Y.~Xiao, M.~Meven,
  R.~Mittal, M.~Rotter, D.~Johrendt, T.~Brueckel, and M.~Loewenhaupt,
\newblock Phys. Rev. B {\bf 79}, 064504 (2009).

\bibitem{YLH+08}
Z.~P. Yin, S.~Leb\`egue, M.~J. Han, B.~P. Neal, S.~Y. Savrasov, and W.~E.
  Pickett,
\newblock Phys. Rev. Lett. {\bf 101}, 047001 (2008).

\bibitem{Sin08a}
D.~J. Singh,
\newblock Phys. Rev. B {\bf 78}, 094511 (2008).

\bibitem{MSJD08}
I.~I. Mazin, D.~J. Singh, M.~D. Johannes, and M.~H. Du,
\newblock Phys. Rev. Lett. {\bf 101}, 057003 (2008).

\bibitem{MJ09}
I.~I. Mazin and M.~D. Johannes,
\newblock Nature Physics {\bf 5}, 141 (2009).

\bibitem{MJB+08}
I.~I. Mazin, M.~D. Johannes, L.~Boeri, K.~Koepernik, and D.~J. Singh,
\newblock Phys. Rev. B {\bf 78}, 085104 (2008).

\bibitem{YLAA09}
A.~N. Yaresko, G.-Q. Liu, V.~N. Antonov, and O.~K. Andersen,
\newblock Phys. Rev. B {\bf 79}, 144421 (2009).

\bibitem{OKZ+09}
I.~Opahle, H.~C. Kandpal, Y.~Zhang, C.~Gros, and R.~Valent\'\i,
\newblock Phys. Rev. B {\bf 79}, 024509 (2009).

\bibitem{FTO+09}
J.~Fink, S.~Thirupathaiah, R.~Ovsyannikov, H.~A. D\"urr, R.~Follath, Y.~Huang,
  S.~de~Jong, M.~S. Golden, Y.-Z. Zhang, H.~O. Jeschke, R.~Valent\'\i,
  C.~Felser, S.~Dastjani~Farahani, M.~Rotter, and D.~Johrendt,
\newblock Phys. Rev. B {\bf 79}, 155118 (2009).

\bibitem{HYPS09}
M.~J. Han, Q.~Yin, W.~E. Pickett, and S.~Y. Savrasov,
\newblock Phys. Rev. Lett. {\bf 102}, 107003 (2009).

\bibitem{YZO+09}
L.~X. Yang, Y.~Zhang, H.~W. Ou, J.~F. Zhao, D.~W. Shen, B.~Zhou, J.~Wei,
  F.~Chen, M.~Xu, C.~He, Y.~Chen, Z.~D. Wang, X.~F. Wang, T.~Wu, G.~Wu, X.~H.
  Chen, M.~Arita, K.~Shimada, M.~Taniguchi, Z.~Y. Lu, T.~Xiang, and D.~L. Feng,
\newblock Phys. Rev. Lett. {\bf 102}, 107002 (2009).

\bibitem{CLLR08}
L.~Craco, M.~S. Laad, S.~Leoni, and H.~Rosner,
\newblock Phys. Rev. B {\bf 78}, 134511 (2008).

\bibitem{JM09}
M.~D. Johannes and I.~I. Mazin,
\newblock Phys. Rev. B {\bf 79}, 220510 (2009).

\bibitem{SBP+11}
A.~Sanna, F.~Bernardini, G.~Profeta, S.~Sharma, J.~K. Dewhurst, A.~Lucarelli,
  L.~Degiorgi, E.~K.~U. Gross, and S.~Massidda,
\newblock Phys. Rev. B {\bf 83}, 054502 (2011).

\bibitem{WCM+12}
P.~Werner, M.~Casula, T.~Miyake, F.~Aryasetiawan, A.~J. Millis, and
  S.~Biermann,
\newblock Nature Physics {\bf 8}, 331 (2012).

\bibitem{LYM+08}
D.~H. Lu, M.~Yi, S.-K. Mo, A.~S. Erickson, J.~Analytis, J.-H. Chu, D.~J. Singh,
  Z.~Hussain, T.~H. Geballe, I.~R. Fisher, and Z.-X. Shen,
\newblock Nature {\bf 455}, 81 (2008).

\bibitem{RTJ+08}
M.~Rotter, M.~Tegel, D.~Johrendt, I.~Schellenberg, W.~Hermes, and R.~P\"ottgen,
\newblock Phys. Rev. B {\bf 78}, 020503 (2008).

\bibitem{GLK+11}
H.~Gretarsson, A.~Lupascu, J.~Kim, D.~Casa, T.~Gog, W.~Wu, S.~R. Julian, Z.~J.
  Xu, J.~S. Wen, G.~D. Gu, R.~H. Yuan, Z.~G. Chen, N.-L. Wang, S.~Khim, K.~H.
  Kim, M.~Ishikado, I.~Jarrige, S.~Shamoto, J.-H. Chu, I.~R. Fisher, and Y.-J.
  Kim,
\newblock Phys. Rev. B {\bf 84}, 100509 (2011).

\bibitem{VFB+12}
P.~Vilmercati, A.~Fedorov, F.~Bondino, F.~Offi, G.~Panaccione, P.~Lacovig,
  L.~Simonelli, M.~A. McGuire, A.~S.~M. Sefat, D.~Mandrus, B.~C. Sales,
  T.~Egami, W.~Ku, and N.~Mannella,
\newblock Phys. Rev. B {\bf 85}, 220503 (2012).

\bibitem{ZIE+09}
V.~B. Zabolotnyy, D.~S. Inosov, D.~V. Evtushinsky, A.~Koitzsch, A.~A. Kordyuk,
  G.~L. Sun, J.~T. Park, D.~Haug, V.~Hinkov, A.~V. Boris, C.~T. Lin,
  M.~Knupfer, A.~N. Yaresko, B.~Buchner, A.~Varykhalov, R.~Follath, and S.~V.
  Borisenko,
\newblock Nature {\bf 457}, 569 (2009).

\bibitem{EIZ+09}
D.~V. Evtushinsky, D.~S. Inosov, V.~B. Zabolotnyy, A.~Koitzsch, M.~Knupfer,
  B.~B\"uchner, M.~S. Viazovska, G.~L. Sun, V.~Hinkov, A.~V. Boris, C.~T. Lin,
  B.~Keimer, A.~Varykhalov, A.~A. Kordyuk, and S.~V. Borisenko,
\newblock Phys. Rev. B {\bf 79}, 054517 (2009).

\bibitem{EIZ+09a}
D.~V. Evtushinsky, D.~S. Inosov, V.~B. Zabolotnyy, M.~S. Viazovska,
  R.~Khasanov, A.~Amato, H.-H. Klauss, H.~Luetkens, C.~Niedermayer, G.~L. Sun,
  V.~Hinkov, C.~T. Lin, A.~Varykhalov, A.~Koitzsch, M.~Knupfer, B.~BÃ¼chner,
  A.~A. Kordyuk, and S.~V. Borisenko,
\newblock New Journal of Physics {\bf 11}, 055069 (2009).

\bibitem{EZK+14}
D.~V. Evtushinsky, V.~B. Zabolotnyy, T.~K. Kim, A.~A. Kordyuk, A.~N. Yaresko,
  J.~Maletz, S.~Aswartham, S.~Wurmehl, A.~V. Boris, D.~L. Sun, C.~T. Lin,
  B.~Shen, H.~H. Wen, A.~Varykhalov, R.~Follath, B.~B\"uchner, and S.~V.
  Borisenko,
\newblock Phys. Rev. B {\bf 89}, 064514 (2014).

\bibitem{YHK11}
Z.~P. Yin, K.~Haule, and G.~Kotliar,
\newblock Nature Physics {\bf 7}, 294 (2011).

\bibitem{FFVJ12}
J.~Ferber, K.~Foyevtsova, R.~Valent\'\i, and H.~O. Jeschke,
\newblock Phys. Rev. B {\bf 85}, 094505 (2012).

\bibitem{YLC+11}
M.~Yi, D.~Lu, J.-H. Chu, J.~G. Analytis, A.~P. Sorini, A.~F. Kemper, B.~Moritz,
  S.-K. Mo, R.~G. Moore, M.~Hashimoto, W.-S. Lee, Z.~Hussain, T.~P. Devereaux,
  I.~R. Fisher, and Z.-X. Shen,
\newblock Proc. Nat. Ac. Sci. US. {\bf 108}, 6878 (2011).

\bibitem{KOK+11}
Y.~Kim, H.~Oh, C.~Kim, D.~Song, W.~Jung, B.~Kim, H.~J. Choi, C.~Kim, B.~Lee,
  S.~Khim, H.~Kim, K.~Kim, J.~Hong, and Y.~Kwon,
\newblock Phys. Rev. B {\bf 83}, 064509 (2011).

\bibitem{CAL+10}
T.-M. Chuang, M.~P. Allan, J.~Lee, Y.~Xie, N.~Ni, S.~L. Bud’ko, G.~S.
  Boebinger, P.~C. Canfield, and J.~C. Davis,
\newblock Science {\bf 327}, 181 (2010).

\bibitem{CAP+10}
J.-H. Chu, J.~G. Analytis, D.~Press, K.~De~Greve, T.~D. Ladd, Y.~Yamamoto, and
  I.~R. Fisher,
\newblock Phys. Rev. B {\bf 81}, 214502 (2010).

\bibitem{ZAY+09}
J.~Zhao, D.~T. Adroja, D.-X. Yao, R.~Bewley, S.~Li, X.~F. Wang, G.~Wu, X.~H.
  Chen, J.~Hu, and P.~Dai,
\newblock Nature Physics {\bf 5}, 555 (2009).

\bibitem{DPM+14}
G.~Derondeau, S.~Polesya, S.~Mankovsky, H.~Ebert, and J.~Min\'ar,
\newblock Phys. Rev. B {\bf 90}, 184509 (2014).

\bibitem{EKM11}
H.~Ebert, D.~K\"odderitzsch, and J.~Min\'{a}r,
\newblock Rep. Prog. Phys. {\bf 74}, 096501 (2011).

\bibitem{SPR-KKR6.3_2}
{H.\ Ebert et al.},
\newblock {\em The Munich SPR-KKR package}, version 6.3,
  http://olymp.cup.uni-muenchen.de/ak/ebert/SPRKKR, 2012.

\bibitem{SJM+08}
A.~S. Sefat, R.~Jin, M.~A. McGuire, B.~C. Sales, D.~J. Singh, and D.~Mandrus,
\newblock Phys. Rev. Lett. {\bf 101}, 117004 (2008).

\bibitem{Veg21}
L.~Vegard,
\newblock Z. Phys. {\bf 5}, 17 (1921).

\bibitem{VWN80}
S.~H. Vosko, L.~Wilk, and M.~Nusair,
\newblock Can. J. Phys. {\bf 58}, 1200 (1980).

\bibitem{MBE13}
J.~Min\'ar, J.~Braun, and H.~Ebert,
\newblock J. Electron. Spectrosc. Relat. Phenom. {\bf 189}, 129 (2013).

\bibitem{BMK+14}
J.~Braun, K.~Miyamoto, A.~Kimura, T.~Okuda, M.~Donath, H.~Ebert, and
  J.~Min\'{a}r,
\newblock New Journal of Physics {\bf 16}, 015005 (2014).

\bibitem{HQB+08}
Q.~Huang, Y.~Qiu, W.~Bao, M.~A. Green, J.~W. Lynn, Y.~C. Gasparovic, T.~Wu,
  G.~Wu, and X.~H. Chen,
\newblock Phys. Rev. Lett. {\bf 101}, 257003 (2008).

\bibitem{RTS+09}
M.~Rotter, M.~Tegel, I.~Schellenberg, F.~M. Schappacher, R.~Pöttgen,
  J.~Deisenhofer, A.~Günther, F.~Schrettle, A.~Loidl, and D.~Johrendt,
\newblock New Journal of Physics {\bf 11}, 025014 (2009).

\bibitem{ABB+08}
A.~A. Aczel, E.~Baggio-Saitovitch, S.~L. Budko, P.~C. Canfield, J.~P. Carlo,
  G.~F. Chen, P.~Dai, T.~Goko, W.~Z. Hu, G.~M. Luke, J.~L. Luo, N.~Ni, D.~R.
  Sanchez-Candela, F.~F. Tafti, N.~L. Wang, T.~J. Williams, W.~Yu, and Y.~J.
  Uemura,
\newblock Phys. Rev. B {\bf 78}, 214503 (2008).

\bibitem{GAB+09}
T.~Goko, A.~A. Aczel, E.~Baggio-Saitovitch, S.~L. Bud'ko, P.~C. Canfield, J.~P.
  Carlo, G.~F. Chen, P.~Dai, A.~C. Hamann, W.~Z. Hu, H.~Kageyama, G.~M. Luke,
  J.~L. Luo, B.~Nachumi, N.~Ni, D.~Reznik, D.~R. Sanchez-Candela, A.~T. Savici,
  K.~J. Sikes, N.~L. Wang, C.~R. Wiebe, T.~J. Williams, T.~Yamamoto, W.~Yu, and
  Y.~J. Uemura,
\newblock Phys. Rev. B {\bf 80}, 024508 (2009).

\bibitem{KOK+09}
D.~Kasinathan, A.~Ormeci, K.~Koch, U.~Burkhardt, W.~Schnelle, A.~Leithe-Jasper,
  and H.~Rosner,
\newblock New Journal of Physics {\bf 11}, 025023 (2009).

\bibitem{AC09}
E.~Akt\"urk and S.~Ciraci,
\newblock Phys. Rev. B {\bf 79}, 184523 (2009).

\bibitem{LCA+09}
C.~Lester, J.-H. Chu, J.~G. Analytis, S.~C. Capelli, A.~S. Erickson, C.~L.
  Condron, M.~F. Toney, I.~R. Fisher, and S.~M. Hayden,
\newblock Phys. Rev. B {\bf 79}, 144523 (2009).

\bibitem{CAK+09}
J.-H. Chu, J.~G. Analytis, C.~Kucharczyk, and I.~R. Fisher,
\newblock Phys. Rev. B {\bf 79}, 014506 (2009).

\bibitem{YLA+09a}
M.~Yi, D.~H. Lu, J.~G. Analytis, J.-H. Chu, S.-K. Mo, R.-H. He, M.~Hashimoto,
  R.~G. Moore, I.~I. Mazin, D.~J. Singh, Z.~Hussain, I.~R. Fisher, and Z.-X.
  Shen,
\newblock Phys. Rev. B {\bf 80}, 174510 (2009).

\bibitem{TKN+09}
M.~A. Tanatar, A.~Kreyssig, S.~Nandi, N.~Ni, S.~L. Bud'ko, P.~C. Canfield,
  A.~I. Goldman, and R.~Prozorov,
\newblock Phys. Rev. B {\bf 79}, 180508 (2009).

\bibitem{KFK+10}
T.~Kondo, R.~M. Fernandes, R.~Khasanov, C.~Liu, A.~D. Palczewski, N.~Ni,
  M.~Shi, A.~Bostwick, E.~Rotenberg, J.~Schmalian, S.~L. Bud'ko, P.~C.
  Canfield, and A.~Kaminski,
\newblock Phys. Rev. B {\bf 81}, 060507 (2010).

\bibitem{EHK80}
D.~E. Eastman, F.~J. Himpsel, and J.~A. Knapp,
\newblock Phys. Rev. Lett. {\bf 44}, 95 (1980).

\bibitem{EP78}
P.~M. Echenique and J.~B. Pendry,
\newblock J. Phys. C: Solid State Phys. {\bf 11}, 2065 (1978).

\bibitem{McR79}
E.~G. McRae,
\newblock Rev. Mod. Phys. {\bf 51}, 541 (1979).

\bibitem{BD02}
J.~Braun and M.~Donath,
\newblock Europhys. Lett. {\bf 59}, 592 (2002).

\bibitem{Hof11}
J.~E. Hoffman,
\newblock Rep. Prog. Phys. {\bf 74}, 124513 (2011).

\bibitem{PFGC10}
G.~Profeta, C.~Franchini, K.~A. I. L.~W. Gamalath, and A.~Continenza,
\newblock Phys. Rev. B {\bf 82}, 195407 (2010).

\bibitem{MJH+09}
F.~Massee, S.~de~Jong, Y.~Huang, J.~Kaas, E.~van Heumen, J.~B. Goedkoop, and
  M.~S. Golden,
\newblock Phys. Rev. B {\bf 80}, 140507 (2009).

\bibitem{NLJ+09}
V.~B. Nascimento, A.~Li, D.~R. Jayasundara, Y.~Xuan, J.~O'Neal, S.~Pan, T.~Y.
  Chien, B.~Hu, X.~B. He, G.~Li, A.~S. Sefat, M.~A. McGuire, B.~C. Sales,
  D.~Mandrus, M.~H. Pan, J.~Zhang, R.~Jin, and E.~W. Plummer,
\newblock Phys. Rev. Lett. {\bf 103}, 076104 (2009).

\end{thebibliography}

\end{document}